\documentclass[12pt]{article}
\usepackage{amssymb}
\renewcommand{\theequation}{\arabic{section}.\arabic{equation}}

\begin{document}

%************************** Text Begins here ******************************

%  Greek letters

\def\a{\alpha}
\def\b{\beta}
\def\d{\delta}
\def\e{\epsilon}
\def\g{\gamma}
\def\h{\mathfrak{h}}
\def\k{\kappa}
\def\l{\lambda}
\def\o{\omega}
\def\p{\wp}
\def\r{\rho}
\def\t{\theta}
\def\s{\sigma}
\def\z{\zeta}
\def\x{\xi}
 \def\A{{\cal{A}}}
 \def\B{{\cal{B}}}
 \def\C{{\cal{C}}}
 \def\D{{\cal{D}}}
 \def\G{{\cal{G}}}
\def\K{{\cal{K}}}
\def\O{\Omega}
\def\R{\bar{R}}
\def\T{{\cal{T}}}
\def\L{\Lambda}
\def\f{E_{\tau,\eta}(sl_2)}
\def\E{E_{\tau,\eta}(sl_n)}
\def\Zb{\mathbb{Z}}
\def\Cb{\mathbb{C}}

\def\R{\overline{R}}
% Shorthands for \begin{equation} and the like

\def\beq{\begin{equation}}
\def\eeq{\end{equation}}
\def\bea{\begin{eqnarray}}
\def\eea{\end{eqnarray}}
\def\ba{\begin{array}}
\def\ea{\end{array}}
\def\no{\nonumber}
\def\le{\langle}
\def\re{\rangle}
\def\lt{\left}
\def\rt{\right}

\newtheorem{Theorem}{Theorem}
\newtheorem{Definition}{Definition}
\newtheorem{Proposition}{Proposition}
\newtheorem{Lemma}{Lemma}
\newtheorem{Corollary}{Corollary}
\newcommand{\proof}[1]{{\bf Proof. }
        #1\begin{flushright}$\Box$\end{flushright}}

\baselineskip=20pt

%%%%%%%%%%%%%%%%%%%%%%%%%%%%%%%%%%%%%%%%%%%%%%%%%%%%%%%%%%%%
%                                                          %
%  Title page                                              %
%                                                          %
%%%%%%%%%%%%%%%%%%%%%%%%%%%%%%%%%%%%%%%%%%%%%%%%%%%%%%%%%%%%
\newfont{\elevenmib}{cmmib10 scaled\magstep1}
\newcommand{\preprint}{
   \begin{flushleft}
     %\elevenmib Yukawa\, Institute\, Kyoto\\
   \end{flushleft}\vspace{-1.3cm}
   \begin{flushright}\normalsize
  % \sf  YITP-03-53\\
    % {\tt hep-th/yymmnnn} \\
    September 2008
   \end{flushright}}
\newcommand{\Title}[1]{{\baselineskip=26pt
   \begin{center} \Large \bf #1 \\ \ \\ \end{center}}}
\newcommand{\Author}{\begin{center}
   \large \bf
Wen-Li Yang ${}^{a,b}$, ~
  Yao-Zhong Zhang ${}^{b}$  and Samuel Kault${}^{b}$
\end{center}}
\newcommand{\Address}{\begin{center}

${}^a$ Institute of Modern Physics, Northwest University,
       Xian 710069, P.R. China\\
${}^b$ The University of Queensland, School of Physical Sciences,
Brisbane,
       QLD 4072, Australia\\

%E-mail: \,wenli@maths.uq.edu.au,\,\, yzz@maths.uq.edu.au

   \end{center}}
\newcommand{\Accepted}[1]{\begin{center}
   {\large \sf #1}\\ \vspace{1mm}{\small \sf Accepted for Publication}
   \end{center}}

\preprint
\thispagestyle{empty}
\bigskip\bigskip\bigskip

\Title{Differential operator realizations of superalgebras and
free field representations of corresponding current algebras}
\Author

\Address
\vspace{1cm}

\begin{abstract}
Based on the particular orderings  introduced for the positive
roots of finite dimensional basic Lie superalgebras, we construct
the explicit differential operator representations of the
$osp(2r|2n)$ and $osp(2r+1|2n)$ superalgebras and the explicit
free field realizations of the corresponding current superalgebras
$osp(2r|2n)_k$ and $osp(2r+1|2n)_k$  at an arbitrary level $k$.
The free field representations of the corresponding
energy-momentum tensors and screening currents of the first kind
are also presented.

\vspace{1truecm} \noindent {\it PACS:} 11.25.Hf; 02.20.Tw

\end{abstract}
\newpage
%%%%%%%%%%%%%%%%%%%%%%%%%%%%%%%%%%%%%%%%%%%%%%%%%%%%%%%%%%%%%%%
%                                                             %
%  1. Introduction                                            %
%                                                             %
%%%%%%%%%%%%%%%%%%%%%%%%%%%%%%%%%%%%%%%%%%%%%%%%%%%%%%%%%%%%%%%
\section{Introduction}
\label{intro} \setcounter{equation}{0}

The interest in two-dimensional non-linear $\sigma$-models with
supergroups or their cosets as target spaces has grown drastically
over the last ten years because of their applications ranging from
string theory \cite{Ber99,Bers99} and logarithmic conformal field
theories (CFTs) \cite{Roz92,Gur93} (for a review, see e.g.
\cite{Flo03,Gab03}, and references therein) to modern condensed
matter physics
\cite{Efe83,Ber95,Mud96,Maa97,Bas00,Gur00,Lud00,Bha01}. The
Wess-Zumino-Novikov-Witten(WZNW) models associated with
supergroups stand out as an important class of such
$\sigma$-models. This is due to the fact that, besides their own
importance, the WZNW models are also the ``building blocks" for
other coset models which can be obtained by gauging or coset
constructions \cite{Met98,Ber00,Kag06,Bab07}. In these models,
current or affine (super)algebras \cite{Kac90} are the underlying
symmetry algebras and are relevant to integrability of the model.

In contrast to the bosonic versions, the WZNW models on
supergroups are far from being {\it understood} (\cite{Sch06} and
references therein), although some progress has been made
\cite{Que07} recently for the models related to type I supergroups
\cite{Kac77}. This is largely due to technical reasons (such as
indecomposability of the operator product expansion (OPE)
\cite{Bel84,Fra97}, appearance of logarithms in correlation
functions and continuous modular transformations of the
irreducible characters \cite{Sem03}), combined with the lack of
``physical intuition".

On the other hand, the Wakimoto free field realizations of current
algebras  \cite{Wak86} have been proved very powerful in the study
of the WZNW models on bosonic groups
\cite{Dos84,Fat86,God86,Ber90,Fur93,And95}. Since the work of
Wakimoto  on the  $sl(2)$ current algebra, much effort has been
made to obtain similar results for the general case
\cite{Fei90,Ber89,Bou90,Ger90,Ito91,Boe97,Ras98}. In these
constructions, the explicit differential operator realizations of
the corresponding finite dimensional (super)algebras play a key
role. However, explicit differential operator expressions heavily
depend on the choice of local coordinate systems in the so-called
big cell ${\cal{U}}$ \cite{Fre06}. Thus it is at least very
involved, if not impossible, to obtain explicit differential
operator expressions for higher-rank (super)algebras in the usual
coordinate systems
\cite{Ger90,Ito91,Bow96,Boe97,Ras98,Din03,Din03-1}. Recently it
was shown in \cite{Yan07,Yan08,Yan08-1} that there exists a
certain coordinate system in ${\cal{U}}$, which drastically
simplifies the computation involved in the construction of
explicit differential operator expressions for  higher-rank
(super)algebras. We call such a coordinate system the ``good
coordinate system".

This paper will show  how to establish a ``good coordinate system"
of the big cell ${\cal{U}}$ for an arbitrary finite-dimensional
basic Lie superalgebra \cite{Kac77}. It will be seen that the
``good coordinate system" {\it indeed} exits and is related to a
particular ordering for the positive roots of the superalgebra.
Based on such an ordering of the positive roots, we construct the
``good coordinate system" for  the superalgebras $osp(2r|2n)$ and
$osp(2r+1|2n)$ and derive their explicit differential operator
representations. We then apply these differential operators to
construct explicit free field representations of the $osp(2r|2n)$
and $osp(2r+1|2n)$ current algebras.

This paper is organized as follows. In section 2, we briefly
review finite-dimensional simple basic Lie superalgebras and their
corresponding current algebras, which also introduces  our
notation and some basic ingredients. In section 3, we introduce
the particular orderings for the positive roots of the
superalgebras $osp(2r|2n)$ and $osp(2r+1|2n)$.  Based  on the
orderings, we construct the explicit differential operator
representations of $osp(2r|2n)$ and $osp(2r+1|2n)$. In section 4
we  apply these differential operator expressions to construct the
explicit free field realizations of  the $osp(2r|2n)$ and
$osp(2r+1|2n)$ currents, the energy-momentum tensors and the
screening currents. Section 5 provides some discussions. In the
Appendix A, we give the matrix forms of the defining
representations of superalgebras $osp(2r|2n)$ and $osp(2r+1|2n)$.

%%%%%%%%%%%%%%%%%%%%%%%%%%%%%%%%%%%%%%%%%%%%%%%%%%%%%%%%%%%%%%%
%                                                             %
%  2. Notation and preliminaries                              %
%                                                             %
%%%%%%%%%%%%%%%%%%%%%%%%%%%%%%%%%%%%%%%%%%%%%%%%%%%%%%%%%%%%%%%

\section{Notation and preliminaries}
\label{CUR} \setcounter{equation}{0}

Let $\G=\G_{\bar{0}}+\G_{\bar{1}}$ be a finite dimensional simple
basic Lie superalgebra \cite{Kac77,Fra00} with a $\Zb_2$-grading:
\bea
 [a]=\lt\{\begin{array}{ll}0&{\rm if}\,a\in\G_{\bar{0}},\\
    1&{\rm if}\,a\in\G_{\bar{1}}.\end{array}\rt.\no
\eea The superdimension of $\G$, denoted by sdim, is defined by
\bea
   {\rm sdim}\lt(\G\rt)={\rm dim}\lt(\G_{\bar{0}}\rt)-{\rm
   dim}\lt(\G_{\bar{1}}\rt).\label{super-dimension}
\eea For any two homogenous elements (i.e. elements with definite
$\Zb_2$-gradings) $a,b \in \G$, the Lie bracket is defined by \bea
 [a,b]=a\,b-(-1)^{[a][b]}b\,a.\no
\eea This (anti)commutator extends to inhomogenous elements
through linearity.  Let $\lt\{E_i|i=1,\ldots,d\rt\}$, where
$d={\rm dim}(\G)$, be the basis of $\G$, which  satisfy
(anti)commutation relations, \bea
 \lt[E_i,\,E_j\rt]=\sum_{l=1}^{d}f_{ij}^l\,E_l.
  \label{construct-constant}
\eea The coefficients $f_{ij}^l$ are the structure constants of
$\G$. Alternatively, one can  use the associated root system
\cite{Fra00} to label the generators of $\G$ as follows. Let $H$
be the Cartan subalgebra of $\G$. A root $\a$ of $\G$ ($\a\neq 0$)
will be  an element in $ H^*$, the dual of $H$, such that: \bea
 \G_{\a}=\lt\{a\in\G|\,[h,a]=\a(h)\,a, \quad \forall h\in H\rt\}\neq
 0.
\eea The set of roots is denoted by $\Delta$. Let $\Pi$:
$=\lt\{\a_i|i=1,\ldots,r\rt\}$ be the simple roots of $\G$, where
the rank of $\G$ is equal to $r={\rm dim}(H)$. With respect to
$\Pi$, the set of positive roots is denoted by $\Delta_+$, and we
write $\a>0$ if $\a\in\Delta_+$. A root $\a$ is called even or
bosonic (odd or fermionic) if $\G_{\a}\in \G_{\bar{0}}$
($\G_{\a}\in \G_{\bar{1}}$). The set of even roots is denoted by
$\Delta_{\bar{0}}$, while the set of odd roots is denoted by
$\Delta_{\bar{1}}$. Associated with each positive root $\a$, there
is a raising operator $E_{\a}$, a lowering operator $F_{\a}$ and a
Cartan generator $H_{\a}$. These operators have definite
$\Zb_2$-gradings: \bea [H_{\a}]=0,\quad
       [E_{\a}]=[F_{\a}]=\lt\{\begin{array}{ll}0,&\a\in\Delta_{\bar{0}}
       \bigcap\Delta_+,\\[2pt]
     1,&\a\in\Delta_{\bar{1}}\bigcap\Delta_+.\end{array} \rt. \no
\eea Moreover, one has the Cartan-Weyl decomposition of $\G$
\bea
 \G=\G_{-}\oplus H \oplus\G_+,\label{Cartan-Weyl}
\eea where $\G_-$ is a span of lowering operators $\{F_{\a}\}$ and
$\G_+$ is a span of raising operators $\{E_{\a}\}$, and $\G_{\pm}$
respectively generates an nilpotent subalgebra of $\G$.

One can introduce  a nondegenerate and invariant supersymmetric
metric or bilinear form  for $\G$, which is  denoted by
$\lt(E_i,E_j\rt)$ (e.g. see (\ref{Bilinear-SD}) for $osp(2r|2n)$
and (\ref{Bilinear-SBC}) for $osp(2r+1|2n)$). Then the affine Lie
superalgebra $\G_k$ (or $\G$ current algebra) associated to $\G$
is generated by $\{E_i^n|i=1,\ldots,d;\,n\in\Zb\}$ satisfying
(anti)commutation relation: \bea
 [E_i^n,\,E_j^m]=\sum_{l=1}^df_{ij}^lE_l^{n+m}+nk(E_i,E_j)\d_{n+m,0}.
 \label{affine-relation}
\eea

Introduce currents
\bea
  E_i(z)=\sum_{n\in\Zb}E_i^n\,z^{-n-1},\quad i=1,\ldots,d.\no
\eea Then the (anti)commutation relations (\ref{affine-relation})
can be re-expressed in terms of the  OPEs \cite{Fra97} of the
currents, \bea
 E_i(z)E_j(w)=k\frac{(E_i,E_j)}{(z-w)^2}
 +\frac{\sum_{m=1}^{d}f_{ij}^lE_l(w)}{(z-w)},\qquad i,j=1,\ldots,d,
 \label{current-OPE}
\eea where $f_{ij}^l$ are the structure constants
(\ref{construct-constant}). The aim of this paper is  to construct
{\it explicit} free field realizations of the current algebras
associated with the unitary series $sl(r|n)$ (or $gl(r|n)$) and
the orthosymplectic series $osp(2r|2n)$ and $osp(2r+1|2n)$ at an
arbitrary level $k$.

%%%%%%%%%%%%%%%%%%%%%%%%%%%%%%%%%%%%%%%%%%%%%%%%%%%%%%%%%%%%%%%%%%%%%%%%%%%%
%                                                                          %
%  3.   Differential operator realizations of superalgebras                %
%                                                                          %
%%%%%%%%%%%%%%%%%%%%%%%%%%%%%%%%%%%%%%%%%%%%%%%%%%%%%%%%%%%%%%%%%%%%%%%%%%%%

\section{Differential operator realizations of superalgebras}
 \label{DIR} \setcounter{equation}{0}

Let $G$ be a Lie supergroup with $\G$ being its Lie superalgebra,
and  $X$ be the flag manifold $G/B_-$, where $B_-$ is the Borel
subgroup corresponding to the subalgebra $\G_-\oplus H$. The
differential operator realization of $\G$ can be obtained from the
infinitesimal action of the corresponding group element on
sections of a line bundle over $X$ \cite{Kos74} or an
$\eta$-invariant lifting of the vector fields on $X$ which form a
representation of $\G$ \cite{Fre06}. As an open set of $X$, we
will take the  big cell ${\cal{U}}$, which is the orbit of the
unit coset under the action of subgroup $N_+$ with Lie
superalgebra $\G_+$. After choosing some local coordinates of
${\cal{U}}$, all the generators of $\G$  in principle can be
realized by first-order differential operators of the coordinates.
In this section we show that there are ``good coordinate systems"
which enable us to obtain the explicit differential operator
realizations of all basic Lie superalgebras. We shall construct
such coordinate systems for the three infinite series of basic
superalgebras $sl(r|n)$, $osp(2r|2n)$ and $osp(2r+1|2n)$ with
generic $r$ and $n$. Our coordinate system in ${\cal{U}}$ is based
on a particular ordering introduced for positive roots
$\Delta_{+}$ of the corresponding superalgebra. We call this
ordering the normal ordering \cite{Kho91} of $\Delta_+$.

\begin{Definition}
The roots of $\Delta_+$ are in normal ordering if all roots are
ordered in such a way that: (i) for any pairwise non-colinear
roots $\a,\b,\g\in \Delta_+$ such that $\g=\a+\b$, $\g$ is between
$\a$ and $\b$; (ii) for $\a,\,2\a\in\Delta_+$, $2\a$ is located on
the nearest right of $\a$.
\end{Definition}

\noindent Such an ordering was constructed explicitly for all
(super)algebras with rank less than 3 in \cite{Kho91}. In the
following, we shall give the normal ordering of positive roots for
each of the three infinite series superalgebras $sl(r|n)$, and
$osp(2r|2n)$ and $osp(2r+1|2n)$.

%%%%%%%%%%%%%%%%%%%%%%%%%%%%%%%%%%%%%%%%%%%%%%%%%%%%%%%%%%%%%
\subsection{Differential operator realization of $sl(r|n)$}
Hereafter, let us fix two non-negative integers $n$ and $r$ such
that $2\leq n+r$. Let us introduce $n+r$ linear-independent
vectors: $\{\d_i|i=1,\ldots,n\}$ and $\{\e_i|i=1,\ldots r\}$.
These vectors are endowed with a symmetric inner product such that
\bea
 (\d_m,\d_l)=\d_{ml},\quad (\d_m,\e_i)=0,\quad
 (\e_i,\e_j)=-\d_{ij} ,\qquad
 i,j=1,\ldots,r,\quad m,l=1,\ldots,n.\label{Inter-product}
\eea The root system $\Delta$ of $sl(r|n)$ (or $A(r-1,n-1)$)  can
be expressed in terms of the vectors:
 \bea
  \Delta=\lt\{\e_i-\e_j,\,\d_{m}-\d_{l},\,
     \d_{m}-\e_i,\,\e_i-\d_{m}\rt\},
     \quad 1\leq i\neq j\leq r,\,\,1\leq m\neq l\leq n,\no
\eea while the even roots $\Delta_{\bar{0}}$ and the odd roots
$\Delta_{\bar{1}}$ are given respectively by
\bea
 \Delta_{\bar{0}}=\lt\{\e_i-\e_j,\,\d_{m}-\d_{l}\rt\},
    \,\,\Delta_{\bar{1}}=\lt\{\pm(\d_{m}-\e_i)\rt\},
    \quad 1\leq i\neq j\leq r,\,\,1\leq m\neq l\leq n.\no
\eea The distinguished simple roots are
 \bea
 &&\a_1=\d_1-\d_2,\ldots,
    \a_{n-1}=\d_{n-1}-\d_n,\,\a_n=\d_{n}-\e_1,\no\\
 &&\a_{n+1}=\e_1-\e_2,\dots,\a_{n+r-1}=\e_{r-1}-\e_r.\no
\eea With regard to the simple roots,  the corresponding positive
roots $\Delta_+$ are \bea
 &&\d_{m}-\d_{l},\quad \e_i-\e_j, \quad
    1\leq i< j\leq r,\,\,1\leq m< l\leq n,\label{positive-roots-SA-1}\\
 &&\d_{m}-\e_i,\quad  1\leq m\leq n,\, 1\leq i\leq r.
   \label{positive-roots-SA-2}
\eea Among these positive roots,
$\{\d_{m}-\e_i|i=1,\ldots,r,\,m=1,\ldots,n\}$ are odd and the
others are even. Then we construct the normal ordering of the
corresponding positive roots.

\begin{Proposition}
A normal ordering  of $\Delta_+$ for $sl(r|n)$ is given by
\bea
   &&\e_{r-1}-\e_{r};\ldots;\,\e_{1}-\e_r,\ldots,\e_1-\e_2;\,\d_n-\e_r,\ldots,
       \d_n-\e_1;\no\\
   &&\quad\ldots;\,\d_1-\e_r,\ldots,\d_1-\e_1,\d_1-\d_n,\ldots,\d_1-\d_2.
       \label{order-SA}
\eea
\end{Proposition}

\vskip0.1in

\noindent {\it Proof}. One can directly verify that the above
ordering of the positive roots
(\ref{positive-roots-SA-1})-(\ref{positive-roots-SA-2}) of
$sl(r|n)$ fulfills all requirements of Definition
1.\hspace{1.truecm}$\Box$

\vskip0.12in

\noindent It is well-known that the big cell ${\cal{U}}$ is
isomorphic to the subgroup $N_+$ and hence to the subalgebra
$\G_+$ via the exponential map. Therefore we can choose the
following coordinate system $G_+(x,\theta)$ for the associated big
cell ${\cal{U}}$: \bea
 G_+(x,\theta)=(G_{n+r-1,n+r})\ldots(G_{j,n+r}\ldots G_{j,j+1})\,
    (G_{1,n+r}\ldots G_{1,2}).\label{Verma-SA-1}
\eea Here, for $i<j$, $G_{i,j}$ is given by \bea
  G_{i,j}=\lt\{\begin{array}{ll} e^{x_{n+i,n+j}E_{\e_i-\e_j}},
           & {\rm if}\,\,\,\, 1\leq i<j\leq r,\\[2pt]
           e^{\theta_{i,n+j}E_{\d_i-\e_j}},& {\rm if}\,\,\,\, 1\leq i\leq n,
           \,\,1\leq j\leq r,\\[2pt]
           e^{x_{i,j}E_{\d_i-\d_j}},& {\rm if}\,\,\,\,
           1\leq i<j\leq n.
           \end{array}\rt.\label{Verma-SA-2}
\eea In the above equations, $\{x_{i,j}\}$ are bosonic coordinates
while $\{\theta_{i,n+j}|1\leq i\leq n,\,1\leq j\leq r\}$ are
fermionic ones. The coordinate system
(\ref{Verma-SA-1})-(\ref{Verma-SA-2}) enabled us \cite{Yan07} to
obtain the explicit differential operator realization of $sl(r|n)$
(or $gl(r|n)$). In the following, we shall show how a similar
normal ordering of the positive roots allows us to construct
``good coordinate systems" in the associated big cell ${\cal{U}}$
of superalgebras $osp(2r|2n)$ and $osp(2r+1|2n)$.

%%%%%%%%%%%%%%%%%%%%%%%%%%%%%%%%%%%%%%%%%%%%%%%%%%%%%%%%%%%%%%%%%%%
\subsection{Differential operator realization of $osp(2r|2n)$}

The root system $\Delta$ of $osp(2r|2n)$ (or $D(r,n)$)  can be
expressed in terms of the vectors $\{\d_l\}$ and $\{\e_i\}$
(\ref{Inter-product}) as follows:
 \bea
  \Delta&=&\lt\{\pm\e_i\pm\e_j,\,\pm\d_m\pm\d_l,\,
     \pm2\d_l,\,\pm\d_l\pm\e_i\rt\},
    \quad 1\leq i\neq j\leq r,\,\,1\leq m\neq l\leq n,\no
\eea while the even roots $\Delta_{\bar{0}}$ and the odd roots
$\Delta_{\bar{1}}$ are given by
\bea
 \Delta_{\bar{0}}&=&\lt\{\pm\e_i\pm\e_j,\,\pm\d_m\pm\d_l,\,\pm2\d_l\rt\},
    \quad \Delta_{\bar{1}}=\lt\{\pm\d_l\pm\e_i\rt\},\no\\
   && \qquad 1\leq i\neq j\leq r,\,\,1\leq m\neq l\leq n.\no
\eea The distinguished simple roots are
 \bea
 &&\a_1=\d_1-\d_2,\ldots,
    \a_{n-1}=\d_{n-1}-\d_n,\,\a_n=\d_{n}-\e_1,\no\\
 &&\a_{n+1}=\e_1-\e_2,\dots,\a_{n+r-1}=\e_{r-1}-\e_r,\,
   \a_{n+r}=\e_{r-1}+\e_{r}.\label{simple-roots-SD(r,n)}
\eea With regard to the simple roots, the corresponding positive
roots $\Delta_+$ are \bea
 &&\d_m-\d_l,\quad 2\d_l,\quad \d_m+\d_l,\quad 1\leq m<l\leq n,
   \label{positive-roots-SD-1}\\
 &&\d_l-\e_i,\quad \d_l+\e_i,\qquad 1\leq i\leq r,\, 1\leq l\leq n,\\
 &&\e_i-\e_j,\quad \e_i+\e_j,\qquad 1\leq i<j\leq r.
   \label{positive-roots-SD-2}
\eea Among these positive roots, $\{\d_l\pm\e_i|i,l=1\ldots,n\}$
are odd and the others are even. Associated with each positive
root $\a$, there is a raising generator $E_{\a}$, a lowering
generator $F_{\a}$ and a Cartan generator $H_{\a}$, giving rise to
the Cartan-Weyl decomposition (\ref{Cartan-Weyl}) of $osp(2r|2n)$:
\bea
   osp(2r|2n)=osp(2r|2n)_-\oplus H_{osp(2r|2n)}\oplus
              osp(2r|2n)_+.
\eea In the defining representation of $osp(2r|2n)$, the matrix
realization of the generators associated with all roots  is given
in Appendix A.1, from which one may derive the structure constants
$f^l_{ij}$ in (\ref{construct-constant}) of the algebra for this
particular choice of the basis.

In order to obtain an explicit differential operator realization
of $osp(2r|2n)$, let us introduce the normal ordering of its
positive roots.
\begin{Proposition}
A normal ordering  of $\Delta_+$ for $osp(2r|2n)$ is given by \bea
   &&\e_{r-1}+\e_{r},\,\e_{r-1}-\e_{r};\ldots;\,\e_{1}+\e_2,\ldots,\e_1+\e_r,\,
       \e_1-\e_r,\ldots,\e_1-\e_2;\no\\
   &&\quad\quad\d_n+\e_1,\ldots,\d_n+\e_r,\,2\d_n,\,\d_n-\e_r,\ldots,
       \d_n-\e_1;\ldots;\no\\
   &&\quad\quad\d_1+\d_2,\ldots,\d_1+\d_n,\,\d_1+\e_1,\ldots,\d_1+\e_r,\,2\d_1,\no\\
   &&\qquad\quad
       \d_1-\e_r,\ldots,\d_1-\e_1,\d_1-\d_n,\ldots,\d_1-\d_2.
       \label{order-SD}
\eea
\end{Proposition}

\vskip0.1in

\noindent {\it Proof}. One can directly verify that the above
ordering of the positive roots
(\ref{positive-roots-SD-1})-(\ref{positive-roots-SD-2}) of
$osp(2r|2n)$ obeys all requirements of Definition
1.\hspace{1.truecm}$\Box$

\vskip0.12in

\noindent For the case  $r=0$, the ordering (\ref{order-SD}) gives
rise to the normal ordering of the positive roots of $sp(2n)$,
while for the case  $n=0$ it yields the normal ordering of the
positive roots of $so(2r)$. Based on these orderings, a ``good
coordinate system" in each of the associated big cells for
$so(2n)$ and $sp(2n)$ was constructed in \cite{Yan08}. Here we use
the ordering (\ref{order-SD}) to construct the ``good coordinate
system" in the associated big cell ${\cal{U}}$ and the explicit
differential operator realization of $osp(2r|2n)$.

%%%%%%%%%%%%%%%%%%%%%%%%%%%%%%%%%%%%%%%%%%%%%%%%%%%%%%%%%%%%%%%%%

Let us introduce a bosonic coordinate ($x_{m,l}$, $\bar{x}_{m,l}$,
$x_l$, $y_{i,j}$ or $\bar{y}_{i,j}$ for $m<l$ and $i<j$) with a
$\Zb_2$-grading zero: $[x]=[\bar{x}]=[y]=[\bar{y}]=0$ associated
with each positive even root (resp. $\d_m-\d_l$, $\d_m+\d_l$,
$2\d_l$, $\e_i-\e_j$ or $\e_i+\e_j$ for $m<l$ and $i<j$), and a
fermionic coordinate ($\theta_{l,i}$ or $\bar{\theta}_{l,i}$) with
a $\Zb_2$-grading one: $[\theta]=[\bar{\theta}]=1$ associated with
each positive odd root (resp. $\d_l-\e_i$ or $\d_l+\e_i$). These
coordinates satisfy the following (anti)commutation relations:
\bea
 &&[x_{i,j},x_{m,l}]=0,\,\,[\partial_{x_{i,j}},\partial_{x_{m,l}}]=0,
   \,\,[\partial_{x_{i,j}},x_{m,l}]=\d_{im}\d_{jl},\label{Fundament-Comm-1}\\
 &&[\bar{x}_{i,j},\bar{x}_{m,l}]=0,\,\,
   [\partial_{\bar{x}_{i,j}},\partial_{\bar{x}_{m,l}}]=0, \,\,
   [\partial_{\bar{x}_{i,j}},\bar{x}_{m,l}]=\d_{im}\d_{jl},\\
 &&[x_m,x_{l}]=0,\,\,[\partial_{x_m},\partial_{x_{l}}]=0,
   \,\,[\partial_{x_{m}},x_{l}]=\d_{ml},\\
 &&[y_{i,j},y_{m,l}]=0,\,\,[\partial_{y_{i,j}},\partial_{y_{m,l}}]=0,
   \,\,[\partial_{y_{i,j}},y_{m,l}]=\d_{im}\d_{jl},\\
 &&[\bar{y}_{i,j},\bar{y}_{m,l}]=0,\,\,
   [\partial_{\bar{y}_{i,j}},\partial_{\bar{y}_{m,l}}]=0, \,\,
   [\partial_{\bar{y}_{i,j}},\bar{y}_{m,l}]=\d_{im}\d_{jl},\\
 &&[\theta_{i,j},\theta_{m,l}]=0,\,\,[\partial_{\theta_{i,j}},\partial_{\theta_{m,l}}]=0,
   \,\,[\partial_{\theta_{i,j}},\theta_{m,l}]=\d_{im}\d_{jl},\\
 &&[\bar{\theta}_{i,j},\bar{\theta}_{m,l}]=0,\,\,
   [\partial_{\bar{\theta}_{i,j}},\partial_{\bar{\theta}_{m,l}}]=0, \,\,
   [\partial_{\bar{\theta}_{i,j}},\bar{\theta}_{m,l}]=\d_{im}\d_{jl},
 \label{Fundament-Comm-2}
\eea and the other (anti)commutation relations vanish.

Based on the very ordering (\ref{order-SD}) of  the positive roots
of $osp(2r|2n)$,  we may introduce  the following coordinate
system $G_+(x,\bar{x};y,\bar{y};\theta,\bar{\theta})$ for the
associated big cell ${\cal{U}}$: \bea
   G_{+}(x,\bar{x};y,\bar{y};\theta,\bar{\theta})&=&
      \lt(\bar{G}_{n+r-1,n+r}\,G_{n+r-1,n+r}\rt)\ldots\no\\
   &&\quad\times  \lt(\bar{G}_{n+1,n+2}\ldots\bar{G}_{n+1,n+r}\,G_{n+1,n+r}
      \ldots G_{n+1,n+2}\rt)\no\\
   &&\quad \times \lt(\bar{G}_{n,n+1}\ldots\bar{G}_{n,n+r}
      \,G_{n}\,G_{n,n+r}\ldots G_{n,n+1}\rt)\ldots\no\\
   &&\quad \times \lt(\bar{G}_{1,2}\ldots\bar{G}_{1,n+r}
      \,G_1\, G_{1,n+r}\ldots G_{1,2}\rt).\label{Coordinate-SD-1}
\eea Here $G_{i,j}$, $\bar{G}_{i,j}$ and $G_i$ are given by \bea
  &&G_{m,l}=e^{x_{m,l}E_{\d_m-\d_l}},\quad
           \bar{G}_{m,l}=e^{\bar{x}_{m,l}E_{\d_m+\d_l}},
           \qquad 1\leq m<l\leq n,\\
  &&G_{l}=e^{x_lE_{2\d_l}},\, G_{l,n+i}=e^{\theta_{l,i}E_{\d_l-\e_i}},\,
           \bar{G}_{l,n+i}=e^{\bar{\theta}_{l,i}E_{\d_l+\e_i}},
           \, 1\leq l\leq n,\,1\leq i\leq r,\\
  &&G_{n+i,n+j}=e^{y_{i,j}E_{\e_i-\e_j}},\qquad
           \bar{G}_{n+i,n+j}=e^{\bar{y}_{i,j}E_{\e_i+\e_j}},
           \qquad 1\leq i<j\leq r.\label{Coordinate-SD-2}
\eea Thus all generators of $osp(2r|2n)$ can be realized  in terms
of the first order differential operators of the coordinates
$\{x,\bar{x};y,\bar{y};\theta,\bar{\theta}\}$ as follows.

Hereafter, let us  adopt the convention that \bea
 E_i\equiv E_{\a_i},\quad F_i\equiv F_{\a_i},
   \quad i=1,\ldots,n+r.\label{Convention}
\eea Let $\langle\L|$ be the highest weight vector of the
representation of $osp(2r|2n)$ with highest weights $\{\l_i\}$ ,
satisfying the following conditions: \bea
 &&\langle\L|F_i=0,\qquad\qquad 1\leq i\leq n+r,\label{highestweight-SD-1}\\
 &&\langle\L|H_i=\l_i\,\langle\L|,\qquad\qquad 1\leq i\leq
  n+r.\label{highestweight-SD-2}
\eea Here the generators $H_i$ are expressed in terms of some
linear combinations of $H_{\a}$ (\ref{SD-H-1})-(\ref{SD-H-2}). An
arbitrary vector in the corresponding Verma module \footnote{The
irreducible highest weight representation can be obtained from the
Verma module through the cohomology procedure \cite{Bou90} with
the help of screening operators (e.g.
(\ref{Scr-P-SD-1})-(\ref{Scr-P-SD-2}) below). } is parametrized by
$\langle\L|$ and the corresponding bosonic and fermionic
coordinates as \bea
 \langle\L;x,\bar{x};y,\bar{y};\theta,\bar{\theta}|=
 \langle\L|G_{+}(x,\bar{x};y,\bar{y};\theta,\bar{\theta}),\label{States-SD}
\eea where $G_{+}(x,\bar{x};y,\bar{y};\theta,\bar{\theta})$ is
given by (\ref{Coordinate-SD-1})-(\ref{Coordinate-SD-2}).

One can define a differential operator realization $\rho^{(d)}$ of
the generators of $osp(2r|2n)$ by \bea
 \rho^{(d)}(g)\,\langle\L;x,\bar{x};y,\bar{y};\theta,\bar{\theta}|
    \equiv \langle\L;x,\bar{x};y,\bar{y};\theta,\bar{\theta}|\,g,\qquad
     \forall g\in osp(2r|2n).\label{definition-SD}
\eea Here $\rho^{(d)}(g)$ is a differential operator of the
coordinates $\{x,\,\bar{x};y,\bar{y};\theta,\bar{\theta}\}$
associated with the generator $g$, which can be obtained from the
defining relation (\ref{definition-SD}). The defining relation
also assures that  the differential operator realization is
actually a representation of $osp(2r|2n)$. Therefore it is
sufficient to give the differential operators related to the
simple roots, as the others can be constructed through the simple
ones by the (anti)commutation relations.  Using the relation
(\ref{definition-SD}) and the Baker-Campbell-Hausdorff formula,
after some algebraic manipulations, we obtain the  differential
operator representation of the simple generators.

\begin{Proposition}
The  differential operator representations of the generators
associated with the simple roots of $osp(2r|2n)$ are given by
 \bea
  \rho^{(d)}(E_l)&=&\sum_{m=1}^{l-1}
    \lt(x_{m,l}\partial_{x_{m,l+1}}-\bar{x}_{m,l+1}\partial_{\bar{x}_{m,l}}\rt)
    +\partial_{x_{l,l+1}},\qquad 1\leq l\leq n-1,\label{Diff-SD-1}\\
  \rho^{(d)}(E_n)&=&\sum_{m=1}^{n-1}
    \lt(x_{m,n}\partial_{\theta_{m,1}}+\bar{\theta}_{m,1}
    \partial_{\bar{x}_{m,n}}\rt)+\partial_{\theta_{n,1}},\\
  \rho^{(d)}(E_{n+i})&=&\sum_{m=1}^n\lt(\theta_{m,i}\partial_{\theta_{m,i+1}}
    -\bar{\theta}_{m,i+1}\partial_{\bar{\theta}_{m,i}}\rt)\no\\
    &&+\sum_{m=1}^{i-1}
    \lt(y_{m,i}\partial_{y_{m,i+1}}-\bar{y}_{m,i+1}\partial_{\bar{y}_{m,i}}\rt)
    +\partial_{y_{i,i+1}},\qquad 1\leq i\leq r-1,\\
  \rho^{(d)}(E_{n+r})&=&\sum_{m=1}^n\lt(2\theta_{m,r-1}\theta_{m,r}\partial_{x_m}+
    \theta_{m,r-1}\partial_{\bar{\theta}_{m,r}}
    -\theta_{m,r}\partial_{\bar{\theta}_{m,r-1}}\rt)\no\\
    &&+\sum_{m=1}^{r-2}
    \lt(y_{m,r-1}\partial_{\bar{y}_{m,r}}-y_{m,r}\partial_{\bar{y}_{m,r-1}}\rt)
    +\partial_{\bar{y}_{r-1,r}},\\[8pt]
  \rho^{(d)}(F_l)&=&\sum_{m=1}^{l-1}\lt(x_{m,l+1}\partial_{x_{m,l}}
    -\bar{x}_{m,l}\partial_{\bar{x}_{m,l+1}}\rt)
    -x_l\partial_{\bar{x}_{l,l+1}}-2\bar{x}_{l,l+1}\partial_{x_{l+1}}\no\\
    &&+\sum_{m=l+2}^n\lt(x_{l,m}\bar{x}_{l,m}\partial_{\bar{x}_{l,l+1}}
    -x_{l,m}\partial_{x_{l+1,m}}
    -2\bar{x}_{l,m}x_{l+1,m}\partial_{x_{l+1}}
    -\bar{x}_{l,m}\partial_{\bar{x}_{l+1,m}}\rt)\no\\
    &&-\sum_{m=1}^r\lt(\theta_{l,m}\bar{\theta}_{l,m}\partial_{\bar{x}_{l,l+1}}
    +\theta_{l,m}\partial_{\theta_{l+1,m}}
    +2\bar{\theta}_{l,m}\theta_{l+1,m}\partial_{x_{l+1}}
    +\bar{\theta}_{l,m}\partial_{\bar{\theta}_{l+1,m}}\rt)\no\\
    &&-x_{l,l+1}^2\partial_{x_{l,l+1}}+2x_{l,l+1}x_{l+1}\partial_{x_{l+1}}
    -2x_{l,l+1}x_l\partial_{x_l}\no\\
    &&+x_{l,l+1}\hspace{-0.1truecm}\lt[
    \sum_{m=l+2}^n
    \hspace{-0.1truecm}\lt(x_{l+1,m}\partial_{x_{l+1,m}}
    \hspace{-0.1truecm}+\hspace{-0.1truecm}
    \bar{x}_{l+1,m}\partial_{\bar{x}_{l+1,m}}
    \hspace{-0.1truecm}-\hspace{-0.1truecm}x_{l,m}\partial_{x_{l,m}}
    -\bar{x}_{l,m}\partial_{\bar{x}_{l,m}}\rt)\rt]\no\\
    &&+x_{l,l+1}\lt[\sum_{m=1}^r\lt(\theta_{l+1,m}\partial_{\theta_{l+1,m}}
    +\bar{\theta}_{l+1,m}\partial_{\bar{\theta}_{l+1,m}}
    -\theta_{l,m}\partial_{\theta_{l,m}}
    -\bar{\theta}_{l,m}\partial_{\bar{\theta}_{l,m}}\rt)\rt]\no\\
    &&+x_{l,l+1}(\l_l-\l_{l+1}),\qquad\qquad 1\leq l\leq
    n-1,\\[6pt]
  \rho^{(d)}(F_n)&=&\sum_{m=1}^{n-1}\lt(\theta_{m,1}\partial_{x_{m,n}}
    -\bar{x}_{m,n}\partial_{\bar{\theta}_{m,1}}\rt)-x_n\partial_{\bar{\theta}_{n,1}}\no\\
    &&+\sum_{m=2}^r\lt(\theta_{n,m}\partial_{y_{1,m}}
    -\theta_{n,m}\bar{\theta}_{n,m}\partial_{\bar{\theta}_{n,1}}
    +\bar{\theta}_{n,m}\partial_{\bar{y}_{1,m}}\rt)\no\\
    &&-\theta_{n,1}\sum_{m=2}^r\lt(\theta_{n,m}\partial_{\theta_{n,m}}
    +\bar{\theta}_{n,m}\partial_{\bar{\theta}_{n,m}}+y_{1,m}\partial_{y_{1,m}}
    +\bar{y}_{1,m}\partial_{\bar{y}_{1,m}}\rt)\no\\
    &&-2\theta_{n,1}x_n\partial_{x_n}
    -2\theta_{n,1}\bar{\theta}_{n,1}\partial_{\bar{\theta}_{n,1}}
    +\theta_{n,1}(\l_{n}+\l_{n+1}),\\
 \rho^{(d)}(F_{n+i})&=&\sum_{m=1}^n(\theta_{m,i+1}\partial_{\theta_{m,i}}
    -\bar{\theta}_{m,i}\partial_{\bar{\theta}_{m,i+1}})
    +\sum_{m=1}^{i-1}(y_{m,i+1}\partial_{y_{m,i}}-\bar{y}_{m,i}\partial_{\bar{y}_{m,i+1}})\no\\
    &&+\sum_{m=i+2}^r\lt(y_{i,m}\bar{y}_{i,m}\partial_{\bar{y}_{i,i+1}}
    -y_{i,m}\partial_{y_{i+1,m}}-\bar{y}_{i,m}\partial_{\bar{y}_{i+1,m}}\rt)\no\\
    &&+y_{i,i+1}\sum_{m=i+2}^r\lt(y_{i+1,m}\partial_{y_{i+1,m}}
    +\bar{y}_{i+1,m}\partial_{\bar{y}_{i+1,m}}
    -y_{i,m}\partial_{y_{i,m}}
    -\bar{y}_{i,m}\partial_{\bar{y}_{i,m}}\rt)\no\\
    &&-y^2_{i,i+1}\partial_{y_{i,i+1}}+y_{i,i+1}(\l_{n+i}-\l_{n+i+1}),
    \qquad 1\leq i\leq r-1,\\
  \rho^{(d)}(F_{n+r})&=&\sum_{m=1}^n\lt(\bar{\theta}_{m,r}\partial_{\theta_{m,r-1}}
    +2\bar{\theta}_{m,r-1}\bar{\theta}_{m,r}\partial_{x_m}
    -\bar{\theta}_{m,r-1}\partial_{\theta_{m,r}}\rt)\no\\
    &&+\sum_{m=1}^{r-2}\lt(\bar{y}_{m,r}\partial_{y_{m,r-1}}-\bar{y}_{m,r-1}\partial_{y_{m,r}} \rt)
    -\bar{y}^2_{r-1,r}\partial_{\bar{y}_{r-1,r}}\no\\
    &&+\bar{y}_{r-1,r}(\l_{n+r-1}+\l_{n+r}),\\[8pt]
  \rho^{(d)}(H_l)&=&\sum_{m=1}^{l-1}\lt(x_{m,l}\partial_{x_{m,l}}
     -\bar{x}_{m,l}\partial_{\bar{x}_{m,l}}\rt)
     -\sum_{m=l+1}^{n}\lt(x_{l,m}\partial_{x_{l,m}}
     +\bar{x}_{l,m}\partial_{\bar{x}_{l,m}}\rt)\no\\
     &&-\sum_{m=1}^r\lt(\theta_{l,m}\partial_{\theta_{l,m}}
     +\bar{\theta}_{l,m}\partial_{\bar{\theta}_{l,m}}\rt)
     -2x_l\partial_{x_l}+\l_l,\qquad 1\leq l\leq n,\\
  \rho^{(d)}(H_{n+i})&=&\sum_{m=1}^n\lt( \theta_{m,i}\partial_{\theta_{m,i}}
     -\bar{\theta}_{m,i}\partial_{\bar{\theta}_{m,i}}\rt)
     +\sum_{m=1}^{i-1}\lt(y_{m,i}\partial_{y_{m,i}}
     -\bar{y}_{m,i}\partial_{\bar{y}_{m,i}}\rt)\no\\
     &&-\sum_{m=i+1}^r\lt(y_{i,m}\partial_{y_{i,m}}
     +\bar{y}_{i,m}\partial_{\bar{y}_{i,m}}\rt)+\l_{n+i},
     \qquad 1\leq i \leq r.\label{Diff-SD-2}
\eea
\end{Proposition}
\vskip0.16in

A direct computation shows that these differential operators
(\ref{Diff-SD-1})-(\ref{Diff-SD-2}) satisfy the $osp(2r|2n)$
(anti)commutation relations corresponding to the simple roots and
the associated Serre relations. This implies that the differential
representation of non-simple generators can be consistently
constructed from the simple ones. Hence, we have obtained an
explicit differential realization of $osp(2r|2n)$.

%%%%%%%%%%%%%%%%%%%%%%%%%%%%%%%%%%%%%%%%%%%%%%%%%%%%%%%%%%%%%%%
\subsection{Differential operator realization of $osp(2r+1|2n)$}

The root system  $\Delta$  of $osp(2r+1|2n)$ (or $B(r,n)$) can be
expressed in terms of the vectors $\{\d_l\}$ and $\{\e_i\}$
(\ref{Inter-product}) as follows:
 \bea
  \Delta=\lt\{\pm\e_i\pm\e_j,\,\pm\e_i,\,\pm\d_m\pm\d_l,\,
     \pm\d_l,\,\pm2\d_l,\,\pm\d_l\pm\e_i\rt\},
     \quad 1\leq i\neq j\leq r,\,\,1\leq m\neq l\leq n,\no
\eea while the even roots $\Delta_{\bar{0}}$ and the odd roots
$\Delta_{\bar{1}}$ are given respectively by
\bea
 \Delta_{\bar{0}}&=&\lt\{\pm\e_i\pm\e_j,\,\pm\e_i,\,\pm\d_m\pm\d_l,\,\pm2\d_l\rt\},
    \quad \Delta_{\bar{1}}=\lt\{\pm\d_l\pm\e_i,\,\pm\d_l\rt\},\no\\
 &&\quad 1\leq i\neq j\leq r,\,\,1\leq m\neq l\leq n.\no
\eea The distinguished simple roots are
 \bea
 &&\a_1=\d_1-\d_2,\ldots,
    \a_{n-1}=\d_{n-1}-\d_n,\,\a_n=\d_{n}-\e_1,\no\\
 &&\a_{n+1}=\e_1-\e_2,\dots,\a_{n+r-1}=\e_{r-1}-\e_r,\,
   \a_{n+r}=\e_{r}.\label{simple-roots-SBC}
\eea With regard to the simple roots,  the corresponding positive
roots $\Delta_+$ are \bea
 &&\d_m-\d_l,\quad 2\d_l,\quad \d_m+\d_l,\quad 1\leq m<l\leq n,\label{positive-roots-SBC-1}\\
 &&\d_l-\e_i,\quad \d_l+\e_i,\quad \d_l,\qquad 1\leq i\leq r,\, 1\leq l\leq n,\\
 &&\e_i-\e_j,\quad \e_i+\e_j,\quad\e_i,\qquad 1\leq i<j\leq r.\label{positive-roots-SBC-2}
\eea Among these positive roots,
$\{\d_l,\,\d_l\pm\e_i|\,i=1,\ldots,r,\,l=1\ldots,n\}$ are odd and
the others are even. Associated with each positive root $\a$,
there is a raising generator $E_{\a}$, a lowering generator
$F_{\a}$ and a Cartan generator $H_{\a}$, giving rise to the
Cartan-Weyl decomposition
 (\ref{Cartan-Weyl}) of $osp(2r+1|2n)$:
\bea
   osp(2r+1|2n)=osp(2r+1|2n)_-\oplus H_{osp(2r+1|2n)}\oplus
              osp(2r+1|2n)_+.
\eea In the defining representation of $osp(2r+1|2n)$, the matrix
realization of the generators associated with all roots  is given
in Appendix A.2, from which one may derive the structure constants
$f^l_{ij}$ in (\ref{construct-constant}) of the algebra for this
particular choice of the basis.

To obtain an explicit expression of the differential operator
realization of $osp(2r+1|2n)$, let us introduce the normal
ordering of its positive roots.
\begin{Proposition}
A normal ordering  of $\Delta_+$ for $osp(2r+1|2n)$ is given by
\bea
   &&\e_r;\,\e_{r-1}+\e_{r},\,\e_{r-1},\,\e_{r-1}-\e_{r};\ldots;\,\e_{1}+\e_2,\ldots,\e_1+\e_r,\,
       \e_1,\,\e_1-\e_r,\ldots,\e_1-\e_2;\no\\
   &&\quad\quad\d_n+\e_1,\ldots,\d_n+\e_r,\,2\d_n,\,\d_n,\,\d_n-\e_r,\ldots,
       \d_n-\e_1;\ldots;\no\\
   &&\qquad\quad\d_1+\d_2,\ldots,\d_1+\d_n,\,\d_1+\e_1,\ldots,\d_1+\e_r,\,2\d_1,\,\d_1,\no\\
   &&\qquad\qquad
       \d_1-\e_r,\ldots,\d_1-\e_1,\d_1-\d_n,\ldots,\d_1-\d_2.
       \label{order-SBC}
\eea
\end{Proposition}

\vskip0.1in

\noindent {\it Proof}. One can directly verify that the above
ordering of the positive roots
(\ref{positive-roots-SBC-1})-(\ref{positive-roots-SBC-2}) of
$osp(2r+1|2n)$ satisfies all requirements of Definition
1.\hspace{1.truecm}$\Box$

\vskip0.12in

\noindent For the case $n=0$, the ordering (\ref{order-SBC}) gives
rise to the normal ordering of the positive roots of $so(2r+1)$.
Based on this ordering  a ``good coordinate system" in the
associated big cell of $so(2r+1)$ was constructed in \cite{Yan08}.
Here we use the ordering (\ref{order-SBC}) to construct a ``good
coordinate system" in the associated big cell ${\cal{U}}$ and the
explicit differential operator realization of $osp(2r+1|2n)$.

%%%%%%%%%%%%%%%%%%%%%%%%%%%%%%%%%%%%%%%%%%%%%%%%%%%%%%%%%%%%%%%%%

In addition to the coordinates $\{x_{m,l},\bar{x}_{m,l};x_m;
y_{i,j},\bar{y}_{i,j}; \theta_{l,i},\bar{\theta}_{l,i}\}$, which
are associated with the positive roots
$\{\d_m-\d_l,\d_m+\d_l;2\d_l;\e_i-\e_j,\e_i+\e_j;
\d_l-\e_i,\d_l+\e_i\}$, we also need to introduce $n+r$ extra
coordinates $\{\theta_l|\,l=1,\ldots,n\}$ and
$\{y_i|\,i=1,\ldots,r\}$ associated with the positive roots
$\{\d_l|\,l=1,\ldots,n\}$ and $\{\e_i|\,i=1,\ldots,r\}$
respectively. The coordinates $\{x_{m,l},\bar{x}_{m,l};x_m;
y_{i,j},\bar{y}_{i,j}; \theta_{l,i},\bar{\theta}_{l,i}\}$ and
their differentials satisfy the same (anti)commutation relations
as (\ref{Fundament-Comm-1})-(\ref{Fundament-Comm-2}). The other
non-trivial relations are \bea
  &&[y_i,y_j]=[\partial_{y_i},\partial_{y_j}]=0,\quad
     [\partial_{y_i},y_j]=\d_{ij},\quad
     i,j=1,\ldots,r.\label{Fundament-Comm-3}\\
  &&[\theta_m,\theta_l]=[\partial_{\theta_m},\partial_{\theta_l}]=0,\quad
    [\partial_{\theta_m},\theta_l]=\d_{ml}, \quad
    m,l=1,\ldots,n.\label{Fundament-Comm-4}
\eea

Based on the very ordering (\ref{order-SBC}) of  the positive
roots of $osp(2r+1|2n)$,  we introduce  the following coordinate
system $G_+(x,\bar{x};y,\bar{y};\theta,\bar{\theta})$ for the
associated big cell ${\cal{U}}$: \bea
   G_{+}(x,\bar{x};y,\bar{y};\theta,\bar{\theta})&=&
      \lt(G_{n+r}\rt)\lt(\bar{G}_{n+r-1,n+r}\,G_{n+r-1}\,G_{n+r-1,n+r}\rt)\ldots\no\\
   &&\quad\times  \lt(\bar{G}_{n+1,n+2}\ldots\bar{G}_{n+1,n+r}\,G_{n+1}\,G_{n+1,n+r}
      \ldots G_{n+1,n+2}\rt)\no\\
   &&\quad \times \lt(\bar{G}_{n,n+1}\ldots\bar{G}_{n,n+r}
      \,\bar{G}_{n}G_{n}\,G_{n,n+r}\ldots G_{n,n+1}\rt)\ldots\no\\
   &&\quad \times \lt(\bar{G}_{1,2}\ldots\bar{G}_{1,n+r}
      \,\bar{G}_1G_1\, G_{1,n+r}\ldots G_{1,2}\rt).\label{Coordinate-SBC-1}
\eea Here $G_{i,j}$,$\bar{G}_{i,j}$,$G_i$ and $\bar{G}_i$ are
given by \bea
  &&G_{m,l}=e^{x_{m,l}E_{\d_m-\d_l}},\quad
           \bar{G}_{m,l}=e^{\bar{x}_{m,l}E_{\d_m+\d_l}},\qquad 1\leq m<l\leq n,\\
  &&\bar{G}_{l}=e^{x_lE_{2\d_l}},\quad G_{l}=e^{\theta_lE_{\d_l}},
           \quad G_{n+i}=e^{y_iE_{\e_i}},
           \quad 1\leq l\leq n, \,1\leq i\leq r,\\
  &&G_{l,n+i}=e^{\theta_{l,i}E_{\d_l-\e_i}},\quad
           \bar{G}_{l,n+i}=e^{\bar{\theta}_{l,i}E_{\d_l+\e_i}},
           \quad 1\leq l\leq n,\,1\leq i\leq r,\\
  &&G_{n+i,n+j}=e^{y_{i,j}E_{\e_i-\e_j}},\qquad
           \bar{G}_{n+i,n+j}=e^{\bar{y}_{i,j}E_{\e_i+\e_j}},\qquad 1\leq i<j\leq r.
           \label{Coordinate-SBC-2}
\eea Then the first order differential operator realization of the
generators of $osp(2r+1|2n)$ can be obtained as follows.

Similarly to the $osp(2r|2n)$ case, we adopt the convention
(\ref{Convention}) for the raising/lowering generators associated
with the simple roots. Let $\langle\L|$ be the highest weight
vector of the representation of $osp(2r+1|2n)$ with highest
weights $\{\l_i\}$ , satisfying the following conditions: \bea
 &&\langle\L|F_i=0,\qquad\qquad 1\leq i\leq n+r,\label{highestweight-SBC-1}\\
 &&\langle\L|H_i=\l_i\,\langle\L|,\qquad\qquad 1\leq i\leq
  n+r.\label{highestweight-SBC-2}
\eea Here the generators $H_i$ are expressed in terms of some
linear combinations of $H_{\a}$ (\ref{SBC-H-1})-(\ref{SBC-H-2}).
An arbitrary vector in the corresponding Verma module is
parametrized by $\langle\L|$ and the corresponding bosonic and
fermionic coordinates as \bea
 \langle\L;x,\bar{x};y,\bar{y};\theta,\bar{\theta}|=
 \langle\L|G_{+}(x,\bar{x};y,\bar{y};\theta,\bar{\theta}),\label{States-SBC}
\eea where $G_{+}(x,\bar{x};y,\bar{y};\theta,\bar{\theta})$ is
given by (\ref{Coordinate-SBC-1})-(\ref{Coordinate-SBC-2})

One can define a differential operator realization $\rho^{(d)}$ of
the generators of $osp(2r+1|2n)$ by \bea
 \rho^{(d)}(g)\,\langle\L;x,\bar{x};y,\bar{y};\theta,\bar{\theta}|
    \equiv \langle\L;x,\bar{x};y,\bar{y};\theta,\bar{\theta}|\,g,\qquad
     \forall g\in osp(2r+1|2n).\label{definition-SBC}
\eea Here $\rho^{(d)}(g)$ is a differential operator of the
coordinates $\{x,\,\bar{x};y,\bar{y};\theta,\bar{\theta}\}$
associated with the generator $g$, which can be obtained from the
defining relation (\ref{definition-SBC}). The defining relation
also assures that  the differential operator realization is
actually a representation of $osp(2r+1|2n)$. Therefore it is
sufficient to give the differential operators related to the
simple roots, as the others can be constructed through the simple
ones by the (anti)commutation relations.

Using the relation (\ref{definition-SBC}) and the
Baker-Campbell-Hausdorff formula, after some algebraic
manipulations, we obtain the following differential operator
representation of the simple generators.

\begin{Proposition} The  differential operator representation of the generators
associated with the simple roots of $osp(2r+1|2n)$ are given by
 \bea
  \rho^{(d)}(E_l)&=&\sum_{m=1}^{l-1}
    \lt(x_{m,l}\partial_{x_{m,l+1}}-\bar{x}_{m,l+1}\partial_{\bar{x}_{m,l}}\rt)
    +\partial_{x_{l,l+1}},\qquad 1\leq l\leq n-1,\label{Diff-SBC-1}\\
  \rho^{(d)}(E_n)&=&\sum_{m=1}^{n-1}
    \lt(x_{m,n}\partial_{\theta_{m,1}}+\bar{\theta}_{m,1}
    \partial_{\bar{x}_{m,n}}\rt)+\partial_{\theta_{n,1}},\\
  \rho^{(d)}(E_{n+i})&=&\sum_{m=1}^n\lt(\theta_{m,i}\partial_{\theta_{m,i+1}}
    -\bar{\theta}_{m,i+1}\partial_{\bar{\theta}_{m,i}}\rt)\no\\
    &&+\sum_{m=1}^{i-1}
    \lt(y_{m,i}\partial_{y_{m,i+1}}-\bar{y}_{m,i+1}\partial_{\bar{y}_{m,i}}\rt)
    +\partial_{y_{i,i+1}},\qquad 1\leq i\leq r-1,\\
  \rho^{(d)}(E_{n+r})&=&\sum_{m=1}^n\lt(\theta_{m}\partial_{\bar{\theta}_{m,r}}-
    \theta_{m,r}\partial_{\theta_{m}}
    -\theta_{m,r}\theta_m\partial_{x_{m}}\rt)\no\\
    &&+\sum_{m=1}^{r-1}
    \lt(y_{m,r}\partial_{y_{m}}-y_{m}\partial_{\bar{y}_{m,r}}\rt)
    +\partial_{y_{r}},\\[8pt]
  \rho^{(d)}(F_l)&=&\sum_{m=1}^{l-1}\lt(x_{m,l+1}\partial_{x_{m,l}}
    -\bar{x}_{m,l}\partial_{\bar{x}_{m,l+1}}\rt)
    -x_l\partial_{\bar{x}_{l,l+1}}-2\bar{x}_{l,l+1}\partial_{x_{l+1}}\no\\
    &&+\sum_{m=l+2}^n\lt(x_{l,m}\bar{x}_{l,m}\partial_{\bar{x}_{l,l+1}}
    -x_{l,m}\partial_{x_{l+1,m}}
    -2\bar{x}_{l,m}x_{l+1,m}\partial_{x_{l+1}}
    -\bar{x}_{l,m}\partial_{\bar{x}_{l+1,m}}\rt)\no\\
    &&-\sum_{m=1}^r\lt(\theta_{l,m}\bar{\theta}_{l,m}\partial_{\bar{x}_{l,l+1}}
    +\theta_{l,m}\partial_{\theta_{l+1,m}}
    +2\bar{\theta}_{l,m}\theta_{l+1,m}\partial_{x_{l+1}}
    +\bar{\theta}_{l,m}\partial_{\bar{\theta}_{l+1,m}}\rt)\no\\
    &&-\theta_l\partial_{\theta_{l+1}}-\theta_l\theta_{l+1}\partial_{x_{l+1}}
    +x_{l,l+1}\theta_{l+1}\partial_{\theta_{l+1}}
    -x_{l,l+1}\theta_{l}\partial_{\theta_{l}}\no\\
    &&-x_{l,l+1}^2\partial_{x_{l,l+1}}+2x_{l,l+1}x_{l+1}\partial_{x_{l+1}}
    -2x_{l,l+1}x_l\partial_{x_l}\no\\
    &&+x_{l,l+1}\hspace{-0.1truecm}\lt[
    \sum_{m=l+2}^n
    \hspace{-0.1truecm}\lt(x_{l+1,m}\partial_{x_{l+1,m}}
    \hspace{-0.1truecm}+\hspace{-0.1truecm}
    \bar{x}_{l+1,m}\partial_{\bar{x}_{l+1,m}}
    \hspace{-0.1truecm}-\hspace{-0.1truecm}x_{l,m}\partial_{x_{l,m}}
    -\bar{x}_{l,m}\partial_{\bar{x}_{l,m}}\rt)\rt]\no\\
    &&+x_{l,l+1}\lt[\sum_{m=1}^r\lt(\theta_{l+1,m}\partial_{\theta_{l+1,m}}
    +\bar{\theta}_{l+1,m}\partial_{\bar{\theta}_{l+1,m}}
    -\theta_{l,m}\partial_{\theta_{l,m}}
    -\bar{\theta}_{l,m}\partial_{\bar{\theta}_{l,m}}\rt)\rt]\no\\
    &&+x_{l,l+1}(\l_l-\l_{l+1}),\qquad\qquad 1\leq l\leq
    n-1,\\[6pt]
  \rho^{(d)}(F_n)&=&\sum_{m=1}^{n-1}\lt(\theta_{m,1}\partial_{x_{m,n}}
    -\bar{x}_{m,n}\partial_{\bar{\theta}_{m,1}}\rt)-x_n\partial_{\bar{\theta}_{n,1}}\no\\
    &&+\sum_{m=2}^r\lt(\theta_{n,m}\partial_{y_{1,m}}
    -\theta_{n,m}\bar{\theta}_{n,m}\partial_{\bar{\theta}_{n,1}}
    +\bar{\theta}_{n,m}\partial_{\bar{y}_{1,m}}\rt)
    -\theta_n\partial_{y_1}\no\\
    &&-\theta_{n,1}\sum_{m=2}^r\lt(\theta_{n,m}\partial_{\theta_{n,m}}
    +\bar{\theta}_{n,m}\partial_{\bar{\theta}_{n,m}}+y_{1,m}\partial_{y_{1,m}}
    +\bar{y}_{1,m}\partial_{\bar{y}_{1,m}}\rt)\no\\
    &&-2\theta_{n,1}x_n\partial_{x_n}
    -2\theta_{n,1}\bar{\theta}_{n,1}\partial_{\bar{\theta}_{n,1}}
    -\theta_{n,1}\theta_n\partial_{\theta_n}-\theta_{n,1}y_1\partial_{y_1}\no\\
    &&+\theta_{n,1}(\l_{n}+\l_{n+1}),\\
 \rho^{(d)}(F_{n+i})&=&\sum_{m=1}^n(\theta_{m,i+1}\partial_{\theta_{m,i}}
    -\bar{\theta}_{m,i}\partial_{\bar{\theta}_{m,i+1}})
    +\sum_{m=1}^{i-1}(y_{m,i+1}\partial_{y_{m,i}}-\bar{y}_{m,i}\partial_{\bar{y}_{m,i+1}})\no\\
    &&+\sum_{m=i+2}^r\lt(y_{i,m}\bar{y}_{i,m}\partial_{\bar{y}_{i,i+1}}
    -y_{i,m}\partial_{y_{i+1,m}}-\bar{y}_{i,m}\partial_{\bar{y}_{i+1,m}}\rt)\no\\
    &&-y_i\partial_{y_{i+1}}+\frac{y_i^2}{2}\partial_{\bar{y}_{i,i+1}}
    +y_{i,i+1}y_{i+1}\partial_{y_{i+1}}-y_{i,i+1}y_{i}\partial_{y_{i}}\no\\
    &&+y_{i,i+1}\sum_{m=i+2}^r\lt(y_{i+1,m}\partial_{y_{i+1,m}}
    +\bar{y}_{i+1,m}\partial_{\bar{y}_{i+1,m}}
    -y_{i,m}\partial_{y_{i,m}}
    -\bar{y}_{i,m}\partial_{\bar{y}_{i,m}}\rt)\no\\
    &&-y^2_{i,i+1}\partial_{y_{i,i+1}}+y_{i,i+1}(\l_{n+i}-\l_{n+i+1}),
    \qquad 1\leq i\leq r-1,\\
  \rho^{(d)}(F_{n+r})&=&\sum_{m=1}^n\lt(\bar{\theta}_{m,r}\partial_{\theta_{m}}
    -\bar{\theta}_{m,r}\theta_{m}\partial_{x_m}
    -\theta_{m}\partial_{\theta_{m,r}}\rt)\no\\
    &&+\sum_{m=1}^{r-1}\lt(y_{m}\partial_{y_{m,r}}-\bar{y}_{m,r}\partial_{y_{m}} \rt)
    -\frac{y^2_{r}}{2}\partial_{y_{r}}+y_{r}\l_{n+r},\\[8pt]
  \rho^{(d)}(H_l)&=&\sum_{m=1}^{l-1}\lt(x_{m,l}\partial_{x_{m,l}}
     -\bar{x}_{m,l}\partial_{\bar{x}_{m,l}}\rt)
     -\sum_{m=l+1}^{n}\lt(x_{l,m}\partial_{x_{l,m}}
     +\bar{x}_{l,m}\partial_{\bar{x}_{l,m}}\rt)\no\\
     &&-\sum_{m=1}^r\lt(\theta_{l,m}\partial_{\theta_{l,m}}
     +\bar{\theta}_{l,m}\partial_{\bar{\theta}_{l,m}}\rt)
     -2x_l\partial_{x_l}-\theta_l\partial_{\theta_l}+\l_l,\quad 1\leq l\leq n,\\
  \rho^{(d)}(H_{n+i})&=&\sum_{m=1}^n\lt( \theta_{m,i}\partial_{\theta_{m,i}}
     -\bar{\theta}_{m,i}\partial_{\bar{\theta}_{m,i}}\rt)
     +\sum_{m=1}^{i-1}\lt(y_{m,i}\partial_{y_{m,i}}
     -\bar{y}_{m,i}\partial_{\bar{y}_{m,i}}\rt)\no\\
     &&-\sum_{m=i+1}^r\lt(y_{i,m}\partial_{y_{i,m}}
     +\bar{y}_{i,m}\partial_{\bar{y}_{i,m}}\rt)-y_i\partial_{y_i}+\l_{n+i},
     \quad 1\leq i \leq r.\label{Diff-SBC-2}
\eea

\end{Proposition}

\vskip0.16in

A direct computation shows that these differential operators
(\ref{Diff-SBC-1})-(\ref{Diff-SBC-2}) satisfy the $osp(2r+1|2n)$
(anti)commutation relations corresponding to the simple roots and
the associated Serre relations. This implies that the differential
representation of non-simple generators can be consistently
constructed from the simple ones. Hence, we have obtained an
explicit differential realization of $osp(2r+1|2n)$.

%%%%%%%%%%%%%%%%%%%%%%%%%%%%%%%%%%%%%%%%%%%%%%%%%%%%%%%%%%%%%%%
%                                                             %
%  4.   Free field realization of current superalgebras       %
%                                                             %
%%%%%%%%%%%%%%%%%%%%%%%%%%%%%%%%%%%%%%%%%%%%%%%%%%%%%%%%%%%%%%%

\section{Free field realization of current superalgebras}
 \label{DIR-1} \setcounter{equation}{0}

%%%%%%%%%%%%%%%%%%%%%%%%%%%%%%%%%%%%%%%%%%%%%%%%%%%%%%%%%%%%%%%%%%%%%
\subsection{Current superalgebra $osp(2r|2n)_k$}

\subsubsection{Free field realization of the currents}
With the help of the explicit differential operator expressions of
$osp(2r|2n)$ given by (\ref{Diff-SD-1})-(\ref{Diff-SD-2}) we can
construct the explicit free field representation of the
$osp(2r|2n)$ current algebra at arbitrary level $k$ in terms of
$n^2+r^2-r$ bosonic $\b$-$\g$ pairs $\{(\b_{i,j},\g_{i,j}),\,
(\bar{\b}_{i,j},\bar{\g}_{i,j}),\,(\b_i,\g_i), $
$(\b'_{i',j'},\g'_{i',j'}),\,
(\bar{\b}'_{i',j'},\bar{\g}'_{i',j'}),\, 1\leq i<j\leq n, \, 1\leq
i'<j'\leq r\}$, $2nr$ fermionic $b-c$ pairs
$\{(\Psi^+_{i,j},\Psi_{i,j}),\,(\bar{\Psi}^+_{i,j},\bar{\Psi}_{i,j}),\,1\leq
i\leq n,\,\, 1\leq j\leq r\}$ and $n+r$ free scalar fields
$\phi_i$, $i=1,\ldots,n+r$. These free fields obey the following
OPEs: \bea
  &&\hspace{-0.8truecm}\b_{i,j}(z)\,\g_{m,l}(w)=-\g_{m,l}(z)\,\b_{i,j}(w)=
      \frac{\d_{im}\d_{jl}}{(z-w)},\,\,1\leq i<j\leq n,\,\,1\leq
      m<l\leq n,\label{OPE-F-1}\\
  &&\hspace{-0.8truecm}\bar{\b}_{i,j}(z)\,\bar{\g}_{m,l}(w)=-\bar{\g}_{m,l}(z)\,
      \bar{\b}_{i,j}(w)= \frac{\d_{im}\d_{jl}}{(z-w)},\,\,1\leq i<j\leq
      n,\,\,1\leq m<l\leq n,\\
  &&\hspace{-0.8truecm}\b_{m}(z)\,\g_{l}(w)=-\g_{m}(z)\,\b_{l}(w)=
      \frac{\d_{ml}}{(z-w)},\,\,1\leq m,l\leq n,\\
  &&\hspace{-0.8truecm}\b'_{i,j}(z)\,\g'_{m,l}(w)=-\g'_{m,l}(z)\,\b'_{i,j}(w)=
      \frac{\d_{im}\d_{jl}}{(z-w)},\,\,1\leq i<j\leq r,\,\,1\leq
      m<l\leq r,\\
  &&\hspace{-0.8truecm}\bar{\b}'_{i,j}(z)\,\bar{\g}'_{m,l}(w)=-\bar{\g}'_{m,l}(z)\,
      \bar{\b}'_{i,j}(w)= \frac{\d_{im}\d_{jl}}{(z-w)},\,\,1\leq i<j\leq
      r,\,\,1\leq m<l\leq r,\\
  &&\hspace{-0.8truecm}\Psi^+_{m,i}(z)\,\Psi_{l,j}(w)=\Psi_{l,j}(z)\,
      \Psi^+_{m,i}(w)= \frac{\d_{ml}\d_{ij}}{(z-w)},\,\,1\leq m,l\leq n,\,\,
      1\leq i,j\leq r,\\
  &&\hspace{-0.8truecm}\bar{\Psi}^+_{m,i}(z)\,\bar{\Psi}_{l,j}(w)=\bar{\Psi}_{l,j}(z)\,
      \bar{\Psi}^+_{m,i}(w)= \frac{\d_{ml}\d_{ij}}{(z-w)},\,\,1\leq m,l\leq n,\,\,
      1\leq i,j\leq r,\\
  &&\hspace{-0.8truecm}\phi_m(z)\phi_l(w)=-\d_{ml}\,
      \ln(z-w),\,\,\,\,\,1\leq m,l\leq n,\\
  &&\hspace{-0.8truecm}\phi_{n+i}(z)\phi_{n+j}(w)=\d_{ij}\,
      \ln(z-w),\,\,\,\,\,1\leq i,j\leq r,\label{OPE-F-2}
\eea and the other OPEs are trivial.

The free field realization of the $osp(2r|2n)$ current algebra is
obtained by the  following substitutions in the differential
operator realization (\ref{Diff-SD-1})-(\ref{Diff-SD-2}) of
$osp(2r|2n)$:
\bea
 &&x_{m,l}\longrightarrow \g_{m,l}(z),\quad \partial_{x_{m,l}}
   \longrightarrow \b_{m,l}(z),\quad 1\leq m<l\leq n,\label{sub-SD-1}\\
 &&\bar{x}_{m,l}\longrightarrow \bar{\g}_{m,l}(z),\quad
   \partial_{\bar{x}_{m,l}} \longrightarrow \bar{\b}_{m,l}(z),
   \quad 1\leq m<l\leq n,\\
 &&x_{l}\longrightarrow \g_{l}(z),\quad \partial_{x_{l}}
   \longrightarrow \b_{l}(z),\quad 1\leq l\leq n,\\
 &&y_{i,j}\longrightarrow \g'_{i,j}(z),\quad \partial_{y_{i,j}}
   \longrightarrow \b'_{i,j}(z),\quad 1\leq i<j\leq r,\\
 &&\bar{y}_{i,j}\longrightarrow \bar{\g}'_{i,j}(z),\quad
   \partial_{\bar{y}_{i,j}} \longrightarrow \bar{\b}'_{i,j}(z),
   \quad 1\leq i<j\leq r,\\
 &&\theta_{l,i}\longrightarrow \Psi^+_{l,i}(z),\quad \partial_{\theta_{l,i}}
   \longrightarrow \Psi_{l,i}(z),\quad 1\leq l\leq n,\,\, 1\leq i\leq r,\\
 &&\bar{\theta}_{l,i}\longrightarrow \bar{\Psi}^+_{l,i}(z),\quad
   \partial_{\bar{\theta}_{l,i}} \longrightarrow \bar{\Psi}_{l,i}(z),
   \quad 1\leq l\leq n,\,\, 1\leq i\leq r,\\
 &&\l_j\longrightarrow \sqrt{k+2(r-n-1)}\,\partial\phi_j(z)\qquad
   1\leq j\leq n+r.\label{sub-SD-2}\eea
Moreover, in order that the resulting free field realization
satisfies the desirable OPEs for $osp(2r|2n)$ currents, one needs
to add certain extra (anomalous) terms which are linear in
$\partial\g(z)$, $\partial\bar{\g}(z)$, $\partial\g'(z)$,
$\partial\bar{\g}'(z)$, $\partial\Psi^+(z)$ and
$\partial\bar{\Psi}^+(z)$ in the expressions of the currents
associated with negative roots (e.g. the last term in the
expressions of $F_i(z)$, see
(\ref{Free-SD-F-1})-(\ref{Free-SD-F-2}) below). Here we present
the results for the currents associated with the simple roots.

\begin{Theorem}
The currents associated with the simple roots of the $osp(2r|2n)$
current algebra at a generic level $k$ are given in terms of the
free fields (\ref{OPE-F-1})-(\ref{OPE-F-2}) as \bea
  E_{l}(z)&=&\sum_{m=1}^{l-1}\lt(\g_{m,l}(z)\b_{m,l+1}(z)-
    \bar{\g}_{m,l+1}(z)\bar{\b}_{m,l}(z)\rt)+\b_{l,l+1}(z),\quad
    1\leq l\leq n-1,\label{Free-SD-1}\\
  E_{n}(z)&=&\sum_{m=1}^{n-1}\lt(\g_{m,n}(z)\Psi_{m,1}(z)+
    \bar{\Psi}^+_{m,1}(z)\bar{\b}_{m,n}(z)\rt)+\Psi_{n,1}(z),\\
  E_{n+i}(z)&=&\sum_{m=1}^n\lt(\Psi^+_{m,i}(z)\Psi_{m,i+1}(z)
    -\bar{\Psi}^+_{m,i+1}(z)\bar{\Psi}_{m,i}(z)\rt)\no\\
    &&+\sum_{m=1}^{i-1}\lt(\g'_{m,i}(z)\b'_{m,i+1}(z)
    \hspace{-0.1truecm}-\hspace{-0.1truecm}
    \bar{\g}'_{m,i+1}(z)\bar{\b}'_{m,i}(z)\rt)
    \hspace{-0.1truecm}+\hspace{-0.1truecm}\b'_{i,i+1}(z),
    \quad 1\leq i\leq r-1,\\
  E_{n+r}(z)&=&\sum_{m=1}^n\hspace{-0.1truecm}
    \lt(2\Psi^+_{m,r-1}(z)\Psi^+_{m,r}(z)
    \b_m(z)\hspace{-0.1truecm}+\hspace{-0.1truecm}
    \Psi^+_{m,r-1}(z)\bar{\Psi}_{m,r}(z)
    \hspace{-0.1truecm}-\hspace{-0.1truecm}
    \Psi^+_{m,r}(z)\bar{\Psi}_{m,r-1}(z)\rt)\no\\
    &&+\sum_{m=1}^{r-2}\lt(\g'_{m,r-1}(z)\bar{\b}'_{m,r}(z)
    -\g'_{m,r}(z)\bar{\b}'_{m,r-1}(z)\rt)+\bar{\b}'_{r-1,r}(z),\\[8pt]
  F_{l}(z)&=&\sum_{m=1}^{l-1}\lt(\g_{m,l+1}(z)\b_{m,l}(z)
    -\bar{\g}_{m,l}(z)\bar{\b}_{m,l+1}(z)\rt)
    -\g_{l}(z)\bar{\b}_{l,l+1}(z)\no\\
    &&-2\bar{\g}_{l,l+1}(z)\b_{l+1}(z)
    +\sum_{m=l+2}^{n}\lt(
    \g_{l,m}(z)\bar{\g}_{l,m}(z)\bar{\b}_{l,l+1}(z)
    -\g_{l,m}(z)\b_{l+1,m}(z)\rt)\no\\
    &&-\sum_{m=l+2}^{n}\lt(
    2\bar{\g}_{l,m}(z)\g_{l+1,m}(z)\b_{l+1}(z)
    +\bar{\g}_{l,m}(z)\bar{\b}_{l+1,m}(z)\rt)\no\\
    &&-\sum_{m=1}^{r}\lt(
    \Psi^+_{l,m}(z)\bar{\Psi}^+_{l,m}(z)\bar{\b}_{l,l+1}(z)
    +\Psi^+_{l,m}(z)\Psi_{l+1,m}(z)\rt)\no\\
    &&-\sum_{m=1}^{r}\lt(
    2\bar{\Psi}^+_{l,m}(z)\Psi^+_{l+1,m}(z)\b_{l+1}(z)
    +\bar{\Psi}^+_{l,m}(z)\bar{\Psi}_{l+1,m}(z)\rt)\no\\
    &&-\g^2_{l,l+1}(z)\b_{l,l+1}(z)-\g_{l,l+1}(z)\sum_{m=l+2}^n
    \lt(\g_{l,m}(z)\b_{l,m}(z)
      +\bar{\g}_{l,m}(z)\bar{\b}_{l,m}(z)\rt)\no\\
    &&+\g_{l,l+1}(z)\sum_{m=l+2}^n\lt(\g_{l+1,m}(z)\b_{l+1,m}(z)+
    \bar{\g}_{l+1,m}(z)\bar{\b}_{l+1,m}(z)\rt)\no\\
    &&-\g_{l,l+1}(z)\sum_{m=1}^r
    \lt(\Psi^+_{l,m}(z)\Psi_{l,m}(z)
      +\bar{\Psi}^+_{l,m}(z)\bar{\Psi}_{l,m}(z)\rt)\no\\
    &&+\g_{l,l+1}(z)\sum_{m=1}^r\lt(\Psi^+_{l+1,m}(z)\Psi_{l+1,m}(z)+
    \bar{\Psi}^+_{l+1,m}(z)\bar{\Psi}_{l+1,m}(z)\rt)\no\\
    &&+2\g_{l,l+1}(z)\g_{l+1}(z)\b_{l+1}(z)
    -2\g_{l,l+1}(z)\g_{l}(z)\b_{l}(z)\no\\
    &&+\sqrt{k+2(r-n-1)}\g_{l,l+1}(z)\lt(\partial\phi_l(z)\hspace{-0.1truecm}
    -\hspace{-0.1truecm}\partial\phi_{l+1}(z)\rt)\no\\
    &&+(-k+2(l-1))\partial\g_{l,l+1}(z),
    \quad 1\leq l\leq n-1,\label{Free-SD-F-1}\\
  F_n(z)&=&\sum_{m=1}^{n-1}\lt(\Psi^+_{m,1}(z)\b_{m,n}(z)
     -\bar{\g}_{m,n}(z)\bar{\Psi}_{m,1}(z)\rt)
     -\g_{n}(z)\bar{\Psi}_{n,1}(z)\no\\
     &&+\sum_{m=2}^r\lt(\Psi^+_{n,m}(z)\b'_{1,m}(z)
     -\Psi^+_{n,m}(z)\bar{\Psi}^+_{n,m}(z)\bar{\Psi}_{n,1}(z)
     +\bar{\Psi}^+_{n,m}(z)\bar{\b}'_{1,m}(z)\rt)\no\\
     &&-\hspace{-0.1truecm}\Psi^+_{n,1}(z)\hspace{-0.1truecm}\sum_{m=2}^r
     \hspace{-0.1truecm}\lt(\Psi^+_{n,m}(z)\Psi_{n,m}(z)
     \hspace{-0.1truecm}+\hspace{-0.1truecm}
     \bar{\Psi}^+_{n,m}(z)\bar{\Psi}_{n,m}(z)\rt)
     \hspace{-0.1truecm}-\hspace{-0.1truecm}
     2\Psi^+_{n,1}(z)\bar{\Psi}^+_{n,1}(z)\bar{\Psi}_{n,1}(z)\no\\
     &&-\Psi^+_{n,1}(z)\sum_{m=2}^r\lt(\g'_{1,m}(z)\b'_{1,m}(z)
     +\bar{\g}'_{1,m}(z)\bar{\b}'_{1,m}(z)\rt)
     -2\Psi^+_{n,1}(z)\g_n(z)\b_n(z)\no\\
     &&+\sqrt{k+2(r-n-1)}\Psi^+_{n,1}(z)\lt(
     \partial\phi_{n}(z)\hspace{-0.1truecm}+\hspace{-0.1truecm}\partial\phi_{n+1}(z)\rt)\no\\
     &&+(-k+2(n-1))\partial\,\Psi^+_{n,1}(z),\\
  F_{n+i}(z)&=&\sum_{m=1}^n\lt(\Psi^+_{m,i+1}(z)\Psi_{m,i}(z)
     -\bar{\Psi}^+_{m,i}(z)\bar{\Psi}_{m,i+1}(z)\rt)\no\\
     &&+\sum_{m=1}^{i-1}\lt(\g'_{m,i+1}(z)\b'_{m,i}(z)
     -\bar{\g}'_{m,i}(z)\bar{\b}'_{m,i+1}(z)\rt)\no\\
     &&+\sum_{m=i+2}^r\lt(\g'_{i,m}(z)\bar{\g}'_{i,m}(z)\bar{\b}'_{i,i+1}(z)
     -\g'_{i,m}(z)\b'_{i+1,m}(z)
     -\bar{\g}'_{i,m}(z)\bar{\b}'_{i+1,m}(z)\rt)\no\\
     &&+\g'_{i,i+1}(z)\sum_{m=i+2}^r\lt(\g'_{i+1,m}(z)\b'_{i+1,m}(z)
     +\bar{\g}'_{i+1,m}(z)\bar{\b}'_{i+1,m}(z)\rt)\no\\
     &&-\g'_{i,i+1}(z)\sum_{m=i+2}^r\lt(\g'_{i,m}(z)\b'_{i,m}(z)
     +\bar{\g}'_{i,m}(z)\bar{\b}'_{i,m}(z)\rt)\no\\
     &&-\g'_{i,i+1}(z)\g'_{i,i+1}(z)\b'_{i,i+1}(z)
     +\sqrt{k+2(r-n-1)}\g'_{i,i+1}(z)\lt(\partial\phi_{n+i}(z)
     -\partial\phi_{n+i+1}(z)\rt)\no\\
     &&+\lt(k+2(i-n-1)\rt)\partial \g'_{i,i+1}(z),\qquad 1\leq i\leq r-1,\\[8pt]
  F_{n+r}(z)&=&\sum_{m=1}^n\hspace{-0.1truecm}\lt(\bar{\Psi}^+_{m,r}(z)\Psi_{m,r-1}(z)
     \hspace{-0.1truecm}+\hspace{-0.1truecm}
     2\bar{\Psi}^+_{m,r-1}(z)\bar{\Psi}^+_{m,r}(z)\b_m(z)
     \hspace{-0.1truecm}-\hspace{-0.1truecm}\bar{\Psi}^+_{m,r-1}(z)\Psi_{m,r}(z)\rt)\no\\
     &&+\sum_{m=1}^{r-2}\lt(\bar{\g}'_{m,r}(z)\b'_{m,r-1}(z)
     -\bar{\g}'_{m,r-1}(z)\b'_{m,r}(z)\rt)\no\\
     &&-\bar{\g}'_{r-1,r}(z)\bar{\g}'_{r-1,r}(z)\bar{\b}'_{r-1,r}(z)\no\\
     &&+\sqrt{k+2(r-n-1)}\bar{\g}'_{r-1,r}(z)\lt(\partial\phi_{n+r-1}(z)
     +\partial\phi_{n+r}(z)\rt)\no\\
     &&+\lt(k+2(r-n-2)\rt)\partial\bar{\g}'_{r-1,r}(z),\label{Free-SD-F-2}\\[8pt]
  H_l(z)&=&\hspace{-0.16truecm}\sum_{m=1}^{l-1}\hspace{-0.16truecm}\lt(\g_{m,l}(z)\b_{m,l}(z)
     \hspace{-0.1truecm}-\hspace{-0.1truecm}\bar{\g}_{m,l}(z)\bar{\b}_{m,l}(z)\rt)
     \hspace{-0.1truecm}-\hspace{-0.1truecm}
     \sum_{m=l+1}^{n}\hspace{-0.16truecm}\lt(\g_{l,m}(z)\b_{l,m}(z)\hspace{-0.1truecm}
     +\hspace{-0.1truecm}\bar{\g}_{l,m}(z)\bar{\b}_{l,m}(z)\rt)\no\\
     &&-2\g_l(z)\b_{l}(z)-\sum_{m=1}^r\lt(\Psi^+_{l,m}(z)\Psi_{l,m}(z)
     +\bar{\Psi}^+_{l,m}(z)\bar{\Psi}_{l,m}(z)\rt)\no\\
     &&+\sqrt{k+2(r-n-1)}\partial\phi_l(z),\qquad\qquad 1\leq l\leq
     n,\\
  H_{n+i}(z)&=&\hspace{-0.16truecm}\sum_{m=1}^{n}\hspace{-0.16truecm}
     \lt(\Psi^+_{m,i}(z)\Psi_{m,i}(z)
     \hspace{-0.1truecm}-\hspace{-0.1truecm}\bar{\Psi}^+_{m,i}(z)\bar{\Psi}_{m,i}(z)\rt)
     \hspace{-0.1truecm}+\hspace{-0.1truecm}
     \sum_{m=1}^{i-1}\hspace{-0.16truecm}\lt(\g'_{m,i}(z)\b'_{m,i}(z)\hspace{-0.1truecm}
     -\hspace{-0.1truecm}\bar{\g}'_{m,i}(z)\bar{\b}'_{m,i}(z)\rt)\no\\
     &&-\sum_{m=i+1}^r\lt(\g'_{i,m}(z)\b'_{i,m}(z)
     +\bar{\g}'_{i,m}(z)\bar{\b}'_{i,m}(z)\rt)\no\\
     &&+\sqrt{k+2(r-n-1)}\partial\phi_{n+i}(z),\qquad\qquad 1\leq i\leq
     r.\label{Free-SD-2}
\eea Here normal ordering of free fields is implied.

\end{Theorem}

\vskip0.1in

\noindent {\it Proof}. It is straightforward to check that the
above free field realization of the currents satisfies the OPEs of
the $osp(2r|2n)$ current algebra: Direct calculation shows that
there are at most second order singularities (e.g.
$1\over(z-w)^{2}$) in the OPEs of the currents. Comparing with the
definition of the current algebra (\ref{current-OPE}), terms with
first order singularity (e.g. the coefficients of $1\over(z-w)$)
are fulfilled due to  the fact that the differential operator
realization (\ref{Diff-SD-1})- (\ref{Diff-SD-2}) is a
representation of the corresponding finite-dimensional
superalgebra $osp(2r|2n)$; terms with second order singularity
$1\over(z-w)^{2}$  also match those in the definition
(\ref{current-OPE}) after the suitable choice we made for the
anomalous terms in the expressions of the currents associated with
negative roots. \hspace{1.truecm}$\Box$

\vskip0.12in

\noindent Some remarks are in order.  The free field realization
of the currents associated with the non-simple roots can be
obtained from the OPEs of the simple ones. For $n=r$, our result
reduces to the free field realization of the $osp(2n|2n)$ current
algebra \cite{Yan08-1}. When $n=0$ (or $r=0$), our result recovers
the free field realization of $so(2r)$ (or $sp(2n)$) current
algebra proposed in \cite{Yan08}.

The free field realization of the $osp(2r|2n)$ current algebra
(\ref{Free-SD-1})-(\ref{Free-SD-2}) gives rise to the Fock
representations of the current algebra in terms of the free fields
(\ref{OPE-F-1})-(\ref{OPE-F-2}). These representations are in
general not irreducible. In order to obtain irreducible ones, one
needs certain screening charges, which are the integrals of
screening currents (see (\ref{Screen-SD-1})-(\ref{Screen-SD-2})
below), and performs the cohomology procedure as in
\cite{Fat86,Fei90,Bou90,Ber89}. We shall construct the associated
screening currents in subsection 4.1.3.

\subsubsection{Energy-momentum tensor} \label{EMT}

In this subsection we construct the free field realization of the
Sugawara energy-momentum tensor $T(z)$ of the $osp(2r|2n)$ current
algebra.  The energy-momentum tensor $T(z)$ can be constructed by
means of the second-order Casimir element of $osp(2r|2n)$, namely,
\bea
  T(z)&=&\frac{1}{2\lt(k+2(r-n-1)\rt)}\lt\{-\sum_{m<l}^n\lt(
     E_{\d_m-\d_l}(z)F_{\d_m-\d_l}(z)+F_{\d_m-\d_l}(z)E_{\d_m-\d_l}(z)\rt)\rt.\no\\
  &&-\sum_{m<l}^n\lt(
     E_{\d_m+\d_l}(z)F_{\d_m+\d_l}(z)+F_{\d_m+\d_l}(z)E_{\d_m+\d_l}(z)\rt)\no\\
  &&-\sum_{l=1}^n\lt[\,2\lt(
     E_{2\d_l}(z)F_{2\d_l}(z)+F_{2\d_l}(z)E_{2\d_l}(z)\rt)+H_l(z)H_l(z)\,\rt]\no\\
  &&+\sum_{l=1}^n\sum_{i=1}^r\lt(
     E_{\d_l-\e_i}(z)F_{\d_l-\e_i}(z)-F_{\d_l-\e_i}(z)E_{\d_l-\e_i}(z)\rt)\no\\
  &&+\sum_{l=1}^n\sum_{i=1}^r\lt(
     E_{\d_l+\e_i}(z)F_{\d_l+\e_i}(z)-F_{\d_l+\e_i}(z)E_{\d_l+\e_i}(z)\rt)\no\\
  &&+\sum_{i<j}^r\lt(E_{\e_i-\e_j}(z)F_{\e_i-\e_j}(z)
     +F_{\e_i-\e_j}(z)E_{\e_i-\e_j}(z)\rt)\no\\
  &&+\lt.\sum_{i<j}^r\lt(E_{\e_i+\e_j}(z)F_{\e_i+\e_j}(z)+F_{\e_i+\e_j}(z)E_{\e_i+\e_j}(z)\rt)
     +\sum_{i=1}^rH_{n+i}(z)H_{n+i}(z)\rt\}.\no
\eea After a tedious calculation, we have

\begin{Proposition} The energy-momentum tensor $T(z)$  of the $osp(2r|2n)$ current
algebra can be expressed in terms of the  free fields
(\ref{OPE-F-1})-(\ref{OPE-F-2}) as \bea T(z)
   &=&-\sum_{l=1}^n\lt(\frac{1}{2}\partial\phi_l(z)\partial\phi_l(z)-
      \frac{n+1-r-l}{\sqrt{k+2(r-n-1)}}\partial^2\phi_l(z)\rt)\no\\
  &&+\sum_{i=1}^r\lt(\frac{1}{2}\partial\phi_{n+i}(z)\partial\phi_{n+i}(z)-
      \frac{r-i}{\sqrt{k+2(r-n-1)}}\partial^2\phi_{n+i}(z)\rt)\no\\
  &&+\sum_{m<l}^n\lt(\b_{m,l}(z)\partial\g_{m,l}(z)+
     \bar{\b}_{m,l}(z)\partial\bar{\g}_{m,l}(z)\rt)
     +\sum_{l=1}^n\b_l(z)\partial\g_l(z)\no\\
  &&+\sum_{i<j}^r\lt(\b'_{i,j}(z)\partial\g'_{i,j}(z)+
     \bar{\b}'_{i,j}(z)\partial\bar{\g}'_{i,j}(z)\rt)\no\\
  &&-\sum_{l=1}^n\sum_{i=1}^r\lt(\Psi_{l,i}(z)\partial\Psi^+_{l,i}(z)
     +\bar{\Psi}_{l,i}(z)\partial\bar{\Psi}^+_{l,i}(z)\rt),\label{Energy-SD}
\eea where normal ordering of free fields is implied. $T(z)$
satisfies  the following OPE:
\bea
   T(z)T(w)=\frac{c/2}{(z-w)^4}+\frac{2T(w)}{(z-w)^2}+\frac{\partial
        T(w)}{(z-w)},
\eea with the central charge $c$ given by
\bea
 c=\frac{(n-r)(2n-2r+1)k}{k+2(r-n-1)}\equiv
 \frac{k\times{\rm sdim}\lt(osp(2r|2n)\rt)}{k+2(r-n-1)}.\label{center-charge-SD}
\eea
\end{Proposition}

\vskip0.16in

It is remarked that for the special case of $n=r$ the central
charge (\ref{center-charge-SD}) vanishes. This  makes the WZNW
model associated with $OSP(2n|2n)$ supergroup  an important class
of CFTs \cite{Ber95,Mud96,Lud00,Bha01,Sch06}. Moreover, we find
that the $osp(2r|2n)$ currents associated with the simple roots
(\ref{Free-SD-1})-(\ref{Free-SD-2}) are indeed primary fields with
conformal dimensional one, namely, \bea
  T(z)E_{i}(w)&=&\frac{E_{i}(w)}{(z-w)^2}+\frac{\partial
    E_{i}(w)}{(z-w)},\,\,1\leq i\leq n+r,\no\\
  T(z)F_{i}(w)&=&\frac{F_{i}(w)}{(z-w)^2}+\frac{\partial
    F_{i}(w)}{(z-w)},\,\,1\leq i\leq n+r,\no\\
  T(z)H_{i}(w)&=&\frac{H_{i}(w)}{(z-w)^2}+\frac{\partial
    H_{i}(w)}{(z-w)},\,\,1\leq i\leq n+r.\no
\eea It is expected that  the $osp(2r|2n)$ currents associated
with non-simple roots, which can be constructed through the simple
ones, are also primary fields with conformal dimensional one.
Therefore, $T(z)$ is the energy-momentum tensor of the
$osp(2r|2n)$ current algebra.

\subsubsection{Screening currents} \label{SC}

Important objects in the application of free field realizations to
the computation of correlation functions  of CFTs are screening
currents. A screening current is a primary field with conformal
dimension one and has the property that the singular part of its
OPE with the affine currents is a total derivative. These
properties ensure that the integrated screening currents
(screening charges) may be inserted into correlators while the
conformal or affine Ward identities remain intact
\cite{Dos84,Ber90}.

Free field realization of screening currents may be constructed
from certain differential operators \cite{Bou90,Ras98} defined by
the relation, \bea
 \rho^{(d)}\lt(s_{\a}\rt)\,
 \langle\L;x,\bar{x};y,\bar{y};\theta,\bar{\theta}|
 \equiv\langle\L|\,E_{\a}\,
 G_{+}(x,\bar{x};y,\bar{y},\theta,\bar{\theta}),\qquad
 {\rm for}\,\,\a\in\Delta_+,
 \label{Def-SD-2}
\eea where $\langle\L|$ is given by
(\ref{highestweight-SD-1})-(\ref{highestweight-SD-2}) and
$G_{+}(x,\bar{x};y,\bar{y},\theta,\bar{\theta})$ is given by
(\ref{Coordinate-SD-1})-(\ref{Coordinate-SD-2}). The operators
$\rho^{(d)}\lt(s_{\a}\rt)$ ($\a\in\Delta_+$) give a differential
operator realization of the subalgebra $osp(2r|2n)_+$. Again it is
sufficient to construct $s_i\equiv \rho^{(d)}\lt(s_{\a_i}\rt)$
related to the simple roots. Using (\ref{Def-SD-2}) and the
Baker-Campbell-Hausdorff formula, after some algebraic
manipulations, we obtain the following explicit expressions for
$s_i$: \bea
   s_{l}&=&\sum_{m=l+2}^n\hspace{-0.08truecm}\lt(
     -\bar{x}_{l+1,m}x_{l+1,m}\partial_{\bar{x}_{l,l+1}}
     \hspace{-0.08truecm}+\hspace{-0.08truecm}
     \bar{x}_{l+1,m}\partial_{\bar{x}_{l,m}}
     \hspace{-0.08truecm}+\hspace{-0.08truecm}
     2x_{l+1,m}\bar{x}_{l,m}\partial_{x_l}
     \hspace{-0.08truecm}+\hspace{-0.08truecm}
     x_{l+1,m}\partial_{x_{l,m}}\rt)\no\\
     &&+\sum_{m=1}^r\hspace{-0.08truecm}\lt(
     -\bar{\theta}_{l+1,m}\theta_{l+1,m}\partial_{\bar{x}_{l,l+1}}
     \hspace{-0.08truecm}+\hspace{-0.08truecm}
     \bar{\theta}_{l+1,m}\partial_{\bar{\theta}_{l,m}}
     \hspace{-0.08truecm}-\hspace{-0.08truecm}
     2\theta_{l+1,m}\bar{\theta}_{l,m}\partial_{x_l}
     \hspace{-0.08truecm}+\hspace{-0.08truecm}
     \theta_{l+1,m}\partial_{\theta_{l,m}}\rt)\no\\
     &&+x_{l+1}\partial_{\bar{x}_{l,l+1}}
     +2\bar{x}_{l,l+1}\partial_{x_l}+\partial_{x_{l,l+1}},
     \qquad 1\leq l\leq n-1,\label{Scr-P-SD-1}\\
   s_n&=&\sum_{m=2}^r\lt(
     \bar{y}_{1,m}\partial_{\bar{\theta}_{n,m}}
     -\bar{y}_{1,m}y_{1,m}\partial_{\bar{\theta}_{n,1}}
     +y_{1,m}\partial_{\theta_{n,m}}
     -2y_{1,m}\bar{\theta}_{n,m}\partial_{x_n}\rt)\no\\
     &&-2\bar{\theta}_{n,1}\partial_{x_n}+\partial_{\theta_{n,1}},\\
   s_{n+i}&=& \sum_{m=i+2}^r\lt(\bar{y}_{i+1,m}\partial_{\bar{y}_{i,m}}
     -\bar{y}_{i+1,m}y_{i+1,m}\partial_{\bar{y}_{i,i+1}}
     +y_{i+1,m}\partial_{y_{i,m}}\rt)\no\\
     &&+\partial_{y_{i,i+1}},\qquad
     1\leq i\leq r-1,\\
   s_{n+r}&=&\partial_{\bar{y}_{r-1,r}}.\label{Scr-P-SD-2}
\eea {}One may obtain the differential operators $s_{\a}$
associated with the non-simple generators from the above simple
ones. Following the procedure similar to \cite{Bou90,Ras98}, we
find the free field realization of the screening currents $S_i(z)$
corresponding to the differential operators $s_i$.

\begin{Proposition} The screening currents associated with the
simple roots of the $osp(2r|2n)$ current algebra at a generic
level $k$ are  given by
 \bea
 S_{l}(z)&=&\lt\{\sum_{m=l+2}^{n}\lt(
    -\bar{\g}_{l+1,m}(z)\g_{l+1,m}(z)\bar{\b}_{l,l+1}(z)
    +\bar{\g}_{l+1,m}(z)\bar{\b}_{l,m}(z)\rt)\rt.\no\\
    &&\quad+\sum_{m=l+2}^{n}\lt(2\g_{l+1,m}(z)\bar{\g}_{l,m}(z)\b_l(z)
    +\g_{l+1,m}(z)\b_{l,m}(z)\rt)+\g_{l+1}(z)\bar{\b}_{l,l+1}(z)\no\\
    &&\quad +2\bar{\g}_{l,l+1}(z)\b_l(z)-\sum_{m=1}^r\lt(
    \bar{\Psi}^+_{l+1,m}(z)\Psi^+_{l+1,m}(z)\bar{\b}_{l,l+1}(z)
    -\bar{\Psi}^+_{l+1,m}(z)\bar{\Psi}_{l,m}(z)\rt)\no\\
    &&\lt.-\sum_{m=1}^r\lt(
    2\Psi^+_{l+1,m}(z)\bar{\Psi}^+_{l,m}(z)\b_{l}(z)
    -\Psi^+_{l+1,m}(z)\Psi_{l,m}(z)\rt)+\b_{l,l+1}(z)\rt\}
    e^{\frac{\a_{l}\cdot\vec{\phi}(z)}{\sqrt{k+2(r-n-1)}}},\no\\
    &&\qquad\qquad 1\leq l\leq n-1,\label{Screen-SD-1}\\[6pt]
 S_{n}(z)&=&\lt\{\sum_{m=2}^r\lt(\bar{\g}'_{1,m}(z)\bar{\Psi}_{n,m}(z)
    -\bar{\g}'_{1,m}(z)\g'_{1,m}(z)\bar{\Psi}_{n,1}(z)
    -2\g'_{1,m}(z)\bar{\Psi}^+_{n,m}(z)\b_n(z)\rt)\rt.\no\\
    &&\quad\lt.+\sum_{m=2}^r\g'_{1,m}(z)\Psi_{n,m}(z)
    -2\bar{\Psi}^+_{n,1}(z)\b_n(z)+\Psi_{n,1}(z)\rt\}
    e^{\frac{\a_{_n}\cdot\vec{\phi}(z)}{\sqrt{k+2(r-n-1)}}},\\
 S_{n+i}(z)&=&\lt\{\sum_{m=i+2}^r\lt(\bar{\g}'_{i+1,m}(z)\bar{\b}'_{i,m}(z)
    -\bar{\g}'_{i+1,m}(z)\g'_{i+1,m}(z)\bar{\b}'_{i,i+1}(z)
    \rt)\rt.\no\\
 &&+\lt.\sum_{m=i+2}^r\g'_{i+1,m}(z)\b'_{i,m}(z)
   +\b'_{i,i+1}(z)\rt\}e^{\frac{\a_{n+i}\cdot\vec{\phi}(z)}{\sqrt{k+2(r-n-1)}}},
    \quad 1\leq i\leq r-1,\\
 S_{n+r}(z)&=&\bar{\b}'_{r-1,r}(z)e^{\frac{\a_{n+r}\cdot\vec{\phi}(z)}{\sqrt{k+2(r-n-1)}}}.
 \label{Screen-SD-2}
\eea Here normal ordering of free fields is implied and
$\vec{\phi}(z)$ is
\bea
 \vec{\phi}(z)=\sum_{i=1}^n\phi_i(z)\,\d_i+\sum_{j=1}^r\phi_{n+j}(z)\,\e_j.\label{Defin-Phi}
\eea
\end{Proposition}

\vskip0.16in

\noindent From a direct calculation, one may find that the OPEs of
the screening currents with the energy-momentum tensor and the
$osp(2r|2n)$ currents (\ref{Free-SD-1})-(\ref{Free-SD-2}) are
\bea
  && T(z)S_i(w)=\frac{S_i(w)}{(z-w)^2}+\frac{\partial
       S_i(w)}{(z-w)}=\partial_w\lt\{\frac{S_i(w)}{(z-w)}\rt\},
       \,\,i=1,\ldots,n+r,\\
 &&E_{i}(z)S_j(w)=0,\qquad i,j=1\ldots,n+r,\\
 &&H_i(z)S_j(w)=0,\qquad i,j=1\ldots,n+r,\\
 &&F_i(z)S_j(w)=(-1)^{[[i]]+[F_i]}\d_{ij}\,
\partial_{w}\lt\{\frac{\lt(k+2(r-n-1)\rt) \,
e^{\frac{\a_i\cdot\vec{\phi}(w)}{\sqrt{k+2(r-n-1)}}}}{(z-w)}\rt\},\no\\
&&\qquad\qquad i,j=1,\ldots,n+r. \eea Here $[[i]]$ is defined  by
\bea
 [[i]]=\lt\{\begin{array}{ll}1,&i=1,\ldots,n,\\
 0,&i=n+1,\ldots,n+r.\end{array}\rt.\label{Def-[]}
\eea The screening currents obtained this way are called screening
currents of the first kind \cite{Ber89}. Moreover, the screening
current $S_n(z)$ is fermionic and the others are bosonic.

%%%%%%%%%%%%%%%%%%%%%%%%%%%%%%%%%%%%%%%%%%%%%%%%%%%%%%%%%%%%%%%%
\subsection{Current superalgebra $osp(2r+1|2n)_k$}

\subsubsection{Free field realization of the currents}

With the help of the explicit differential operator expressions of
$osp(2r+1|2n)$ given by (\ref{Diff-SBC-1})-(\ref{Diff-SBC-2}) we
can construct the explicit free field representation of the
$osp(2r+1|2n)$ current algebra at an arbitrary level $k$ in terms
of $n^2+r^2$ bosonic $\b$-$\g$ pairs $\{(\b_{i,j},\g_{i,j}),\,
(\bar{\b}_{i,j},\bar{\g}_{i,j}),\,(\b_i,\g_i),$
$(\b'_{i',j'},\g'_{i',j'}),\,
(\bar{\b}'_{i',j'},\bar{\g}'_{i',j'}),$ $(\b'_{i'},\g'_{i'}),\,
1\leq i<j\leq n, \, 1\leq i'<j'\leq r\}$, $n(2r+1)$ fermionic
$b-c$ pairs
$\{(\Psi^+_{i,j},\Psi_{i,j}),\,(\bar{\Psi}^+_{i,j},\bar{\Psi}_{i,j}),$
$(\Psi^+_i,\Psi_i),\,1\leq i\leq n,\,\, 1\leq j\leq r\}$ and $n+r$
free scalar fields $\phi_i$, $i=1,\ldots,n+r$. The free fields
$\{(\b_{i,j},\g_{i,j}),\,
(\bar{\b}_{i,j},\bar{\g}_{i,j}),\,(\b_i,\g_i),\,(\b'_{i',j'},\g'_{i',j'}),\,
(\bar{\b}'_{i',j'},\bar{\g}'_{i',j'})\}$,
$\{(\Psi^+_{i,j},\Psi_{i,j}),\,(\bar{\Psi}^+_{i,j},\bar{\Psi}_{i,j})\}$
and $\{\phi_i\}$ obey the same OPEs as
(\ref{OPE-F-1})-(\ref{OPE-F-2}). The other non-trivial OPEs are
\bea
 &&\hspace{-0.8truecm}\b'_{i}(z)\,\g'_{j}(w)=-\g'_{j}(z)\,\b'_{i}(w)=
      \frac{\d_{ij}}{(z-w)},\,\,\quad 1\leq i,j\leq r,\label{OPE-F-3}\\
 &&\hspace{-0.8truecm}\Psi^+_{m}(z)\,\Psi_{l}(w)=\Psi_{l}(z)\,\Psi^+_{m}(w)=
      \frac{\d_{ml}}{(z-w)},\,\,\quad 1\leq m,l\leq n.\label{OPE-F-4}
\eea

The free field realization of the $osp(2r+1|2n)$ current algebra
is obtained by the following substitutions in the differential
operator realization (\ref{Diff-SBC-1})-(\ref{Diff-SBC-2}) of
$osp(2r+1|2n)$:
\bea
 &&x_{m,l}\longrightarrow \g_{m,l}(z),\quad \partial_{x_{m,l}}
   \longrightarrow \b_{m,l}(z),\quad 1\leq m<l\leq n,\label{sub-SBC-1}\\
 &&\bar{x}_{m,l}\longrightarrow \bar{\g}_{m,l}(z),\quad
   \partial_{\bar{x}_{m,l}} \longrightarrow \bar{\b}_{m,l}(z),
   \quad 1\leq m<l\leq n,\\
 &&x_{l}\longrightarrow \g_{l}(z),\quad \partial_{x_{l}}
   \longrightarrow \b_{l}(z),\quad 1\leq l\leq n,\\
 &&y_{i,j}\longrightarrow \g'_{i,j}(z),\quad \partial_{y_{i,j}}
   \longrightarrow \b'_{i,j}(z),\quad 1\leq i<j\leq r,\\
 &&\bar{y}_{i,j}\longrightarrow \bar{\g}'_{i,j}(z),\quad
   \partial_{\bar{y}_{i,j}} \longrightarrow \bar{\b}'_{i,j}(z),
   \quad 1\leq i<j\leq r,\\
 &&y_i\longrightarrow \g'_{i}(z),\quad \partial_{y_i}\longrightarrow \b'_{i}(z),
   \quad 1\leq i\leq r,\\
 &&\theta_{l,i}\longrightarrow \Psi^+_{l,i}(z),\quad \partial_{\theta_{l,i}}
   \longrightarrow \Psi_{l,i}(z),\quad 1\leq l\leq n,\,\, 1\leq i\leq r,\\
 &&\bar{\theta}_{l,i}\longrightarrow \bar{\Psi}^+_{l,i}(z),\quad
   \partial_{\bar{\theta}_{l,i}} \longrightarrow \bar{\Psi}_{l,i}(z),
   \quad 1\leq l\leq n,\,\, 1\leq i\leq r,\\
 &&\theta_l\longrightarrow\Psi^+_l(z),
   \quad \partial_{\theta_l}\longrightarrow\Psi_l(z),
   \quad 1\leq l\leq n,\\
 &&\l_j\longrightarrow \sqrt{k+2r-2n-1}\,\partial\phi_j(z)\qquad
   1\leq j\leq n+r,\label{sub-SBC-2}\eea
followed with by the addition of anomalous terms linear in
$\partial\g(z)$, $\partial\bar{\g}(z)$, $\partial\g'(z)$,
$\partial\bar{\g}'(z)$, $\partial\Psi^+(z)$ and
$\partial\bar{\Psi}^+(z)$ in the expressions of the currents. Here
we present the results for the currents associated with the simple
roots.
\begin{Theorem} The currents associated with the simple roots
of the $osp(2r+1|2n)$ current algebra at a generic level $k$ are
given in terms of the free fields (\ref{OPE-F-1})-(\ref{OPE-F-2})
and (\ref{OPE-F-3})-(\ref{OPE-F-4}) as
 \bea
  E_{l}(z)&=&\sum_{m=1}^{l-1}\lt(\g_{m,l}(z)\b_{m,l+1}(z)-
    \bar{\g}_{m,l+1}(z)\bar{\b}_{m,l}(z)\rt)+\b_{l,l+1}(z),\quad
    1\leq l\leq n-1,\label{Free-SBC-1}\\
  E_{n}(z)&=&\sum_{m=1}^{n-1}\lt(\g_{m,n}(z)\Psi_{m,1}(z)+
    \bar{\Psi}^+_{m,1}(z)\bar{\b}_{m,n}(z)\rt)+\Psi_{n,1}(z),\\
  E_{n+i}(z)&=&\sum_{m=1}^n\lt(\Psi^+_{m,i}(z)\Psi_{m,i+1}(z)
    -\bar{\Psi}^+_{m,i+1}(z)\bar{\Psi}_{m,i}(z)\rt)\no\\
    &&+\sum_{m=1}^{i-1}\lt(\g'_{m,i}(z)\b'_{m,i+1}(z)
    \hspace{-0.1truecm}-\hspace{-0.1truecm}
    \bar{\g}'_{m,i+1}(z)\bar{\b}'_{m,i}(z)\rt)
    \hspace{-0.1truecm}+\hspace{-0.1truecm}\b'_{i,i+1}(z),
    \quad 1\leq i\leq r-1,\\
  E_{n+r}(z)&=&\sum_{m=1}^n\hspace{-0.1truecm}
    \lt(\Psi^+_{m}(z)\bar{\Psi}_{m,r}(z)
    \hspace{-0.1truecm}-\hspace{-0.1truecm}
    \Psi^+_{m,r}(z)\Psi_{m}(z)
    \hspace{-0.1truecm}-\hspace{-0.1truecm}
    \Psi^+_{m,r}(z)\Psi^+_{m}(z)\b_m(z)\rt)\no\\
    &&+\sum_{m=1}^{r-1}\lt(\g'_{m,r}(z)\b'_{m}(z)
    -\g'_{m}(z)\bar{\b}'_{m,r}(z)\rt)+\b'_{r}(z),\\[8pt]
  F_{l}(z)&=&\sum_{m=1}^{l-1}\lt(\g_{m,l+1}(z)\b_{m,l}(z)
    -\bar{\g}_{m,l}(z)\bar{\b}_{m,l+1}(z)\rt)
    -\g_{l}(z)\bar{\b}_{l,l+1}(z)\no\\
    &&-2\bar{\g}_{l,l+1}(z)\b_{l+1}(z)
    +\sum_{m=l+2}^{n}\lt(
    \g_{l,m}(z)\bar{\g}_{l,m}(z)\bar{\b}_{l,l+1}(z)
    -\g_{l,m}(z)\b_{l+1,m}(z)\rt)\no\\
    &&-\sum_{m=l+2}^{n}\lt(
    2\bar{\g}_{l,m}(z)\g_{l+1,m}(z)\b_{l+1}(z)
    +\bar{\g}_{l,m}(z)\bar{\b}_{l+1,m}(z)\rt)\no\\
    &&-\sum_{m=1}^{r}\lt(
    \Psi^+_{l,m}(z)\bar{\Psi}^+_{l,m}(z)\bar{\b}_{l,l+1}(z)
    +\Psi^+_{l,m}(z)\Psi_{l+1,m}(z)\rt)\no\\
    &&-\sum_{m=1}^{r}\lt(
    2\bar{\Psi}^+_{l,m}(z)\Psi^+_{l+1,m}(z)\b_{l+1}(z)
    +\bar{\Psi}^+_{l,m}(z)\bar{\Psi}_{l+1,m}(z)\rt)\no\\
    &&-\Psi^+_l(z)\Psi_{l+1}(z)-\Psi^+_l(z)\Psi^+_{l+1}(z)\b_{l+1}(z)\no\\
    &&-\g^2_{l,l+1}(z)\b_{l,l+1}(z)-\g_{l,l+1}(z)\sum_{m=l+2}^n
    \lt(\g_{l,m}(z)\b_{l,m}(z)
      +\bar{\g}_{l,m}(z)\bar{\b}_{l,m}(z)\rt)\no\\
    &&+\g_{l,l+1}(z)\sum_{m=l+2}^n\lt(\g_{l+1,m}(z)\b_{l+1,m}(z)+
    \bar{\g}_{l+1,m}(z)\bar{\b}_{l+1,m}(z)\rt)\no\\
    &&-\g_{l,l+1}(z)\sum_{m=1}^r
    \lt(\Psi^+_{l,m}(z)\Psi_{l,m}(z)
      +\bar{\Psi}^+_{l,m}(z)\bar{\Psi}_{l,m}(z)\rt)\no\\
    &&+\g_{l,l+1}(z)\sum_{m=1}^r\lt(\Psi^+_{l+1,m}(z)\Psi_{l+1,m}(z)+
    \bar{\Psi}^+_{l+1,m}(z)\bar{\Psi}_{l+1,m}(z)\rt)\no\\
    &&+2\g_{l,l+1}(z)\g_{l+1}(z)\b_{l+1}(z)
    -2\g_{l,l+1}(z)\g_{l}(z)\b_{l}(z)\no\\
    &&+\g_{l,l+1}(z)\Psi^+_{l+1}(z)\Psi_{l+1}(z)-\g_{l,l+1}(z)\Psi^+_{l}(z)\Psi_{l}(z)\no\\
    &&+\sqrt{k+2r-2n-1}\g_{l,l+1}(z)\lt(\partial\phi_l(z)\hspace{-0.1truecm}
    -\hspace{-0.1truecm}\partial\phi_{l+1}(z)\rt)\no\\
    &&+(-k+2(l-1))\partial\g_{l,l+1}(z),
    \quad 1\leq l\leq n-1,\label{Free-SBC-F-1}\\
  F_n(z)&=&\sum_{m=1}^{n-1}\lt(\Psi^+_{m,1}(z)\b_{m,n}(z)
     -\bar{\g}_{m,n}(z)\bar{\Psi}_{m,1}(z)\rt)
     -\g_{n}(z)\bar{\Psi}_{n,1}(z)\no\\
     &&+\sum_{m=2}^r\lt(\Psi^+_{n,m}(z)\b'_{1,m}(z)
     -\Psi^+_{n,m}(z)\bar{\Psi}^+_{n,m}(z)\bar{\Psi}_{n,1}(z)
     +\bar{\Psi}^+_{n,m}(z)\bar{\b}'_{1,m}(z)\rt)\no\\
     &&-\hspace{-0.1truecm}\Psi^+_{n,1}(z)\hspace{-0.1truecm}\sum_{m=2}^r
     \hspace{-0.1truecm}\lt(\Psi^+_{n,m}(z)\Psi_{n,m}(z)
     \hspace{-0.1truecm}+\hspace{-0.1truecm}
     \bar{\Psi}^+_{n,m}(z)\bar{\Psi}_{n,m}(z)\rt)
     \hspace{-0.1truecm}-\hspace{-0.1truecm}
     2\Psi^+_{n,1}(z)\bar{\Psi}^+_{n,1}(z)\bar{\Psi}_{n,1}(z)\no\\
     &&-\Psi^+_{n,1}(z)\sum_{m=2}^r\lt(\g'_{1,m}(z)\b'_{1,m}(z)
     +\bar{\g}'_{1,m}(z)\bar{\b}'_{1,m}(z)\rt)
     -2\Psi^+_{n,1}(z)\g_n(z)\b_n(z)\no\\
     &&-\Psi^+_n(z)\b'_1(z)-\Psi^+_{n,1}(z)\Psi^+_n(z)\Psi_n(z)
     -\Psi^+_{n,1}(z)\g'_1(z)\b'_1(z)\no\\
     &&+\sqrt{k+2r-2n-1}\Psi^+_{n,1}(z)\lt(
     \partial\phi_{n}(z)\hspace{-0.1truecm}+\hspace{-0.1truecm}\partial\phi_{n+1}(z)\rt)\no\\
     &&+(-k+2(n-1))\partial\,\Psi^+_{n,1}(z),\\
  F_{n+i}(z)&=&\sum_{m=1}^n\lt(\Psi^+_{m,i+1}(z)\Psi_{m,i}(z)
     -\bar{\Psi}^+_{m,i}(z)\bar{\Psi}_{m,i+1}(z)\rt)\no\\
     &&+\sum_{m=1}^{i-1}\lt(\g'_{m,i+1}(z)\b'_{m,i}(z)
     -\bar{\g}'_{m,i}(z)\bar{\b}'_{m,i+1}(z)\rt)\no\\
     &&+\sum_{m=i+2}^r\lt(\g'_{i,m}(z)\bar{\g}'_{i,m}(z)\bar{\b}'_{i,i+1}(z)
     -\g'_{i,m}(z)\b'_{i+1,m}(z)
     -\bar{\g}'_{i,m}(z)\bar{\b}'_{i+1,m}(z)\rt)\no\\
     &&-\g'_i(z)\b'_{i+1}(z)+\frac{1}{2}\g'_i(z)\g'_i(z)\bar{\b}'_{i,i+1}(z)\no\\
     &&+\g'_{i,i+1}(z)\sum_{m=i+2}^r\lt(\g'_{i+1,m}(z)\b'_{i+1,m}(z)
     +\bar{\g}'_{i+1,m}(z)\bar{\b}'_{i+1,m}(z)\rt)\no\\
     &&-\g'_{i,i+1}(z)\sum_{m=i+2}^r\lt(\g'_{i,m}(z)\b'_{i,m}(z)
     +\bar{\g}'_{i,m}(z)\bar{\b}'_{i,m}(z)\rt)\no\\
     &&-\g'_{i,i+1}(z)\lt(\g'_{i,i+1}(z)\b'_{i,i+1}(z)
     -\g'_{i+1}(z)\b'_{i+1}(z)+\g'_{i}(z)\b'_{i}(z)\rt)\no\\
     &&+\sqrt{k+2r-2n-1}\g'_{i,i+1}(z)\lt(\partial\phi_{n+i}(z)
     -\partial\phi_{n+i+1}(z)\rt)\no\\
     &&+\lt(k+2(i-n-1)\rt)\partial \g'_{i,i+1}(z),\qquad 1\leq i\leq r-1,\\[8pt]
  F_{n+r}(z)&=&\sum_{m=1}^n\hspace{-0.1truecm}
     \lt(\bar{\Psi}^+_{m,r}(z)\Psi_{m}(z)
     \hspace{-0.1truecm}-\hspace{-0.1truecm}
     \bar{\Psi}^+_{m,r}(z)\Psi^+_{m}(z)\b_m(z)
     \hspace{-0.1truecm}-\hspace{-0.1truecm}\Psi^+_{m}(z)\Psi_{m,r}(z)\rt)\no\\
     &&+\sum_{m=1}^{r-1}\lt(\g'_{m}(z)\b'_{m,r}(z)
     -\bar{\g}'_{m,r}(z)\b'_{m}(z)\rt)-\frac{1}{2}\g'_{r}(z)\g'_{r}(z)\b'_{r}(z)\no\\
     &&+\sqrt{k+2r-2n-1}\g'_{r}(z)\partial\phi_{n+r}(z)
     +\lt(k+2(r-n-1)\rt)\partial\g'_{r}(z),\label{Free-SBC-F-2}\\[8pt]
  H_l(z)&=&\hspace{-0.16truecm}\sum_{m=1}^{l-1}\hspace{-0.16truecm}\lt(\g_{m,l}(z)\b_{m,l}(z)
     \hspace{-0.1truecm}-\hspace{-0.1truecm}\bar{\g}_{m,l}(z)\bar{\b}_{m,l}(z)\rt)
     \hspace{-0.1truecm}-\hspace{-0.1truecm}
     \sum_{m=l+1}^{n}\hspace{-0.16truecm}\lt(\g_{l,m}(z)\b_{l,m}(z)\hspace{-0.1truecm}
     +\hspace{-0.1truecm}\bar{\g}_{l,m}(z)\bar{\b}_{l,m}(z)\rt)\no\\
     &&-2\g_l(z)\b_{l}(z)-\sum_{m=1}^r\lt(\Psi^+_{l,m}(z)\Psi_{l,m}(z)
     +\bar{\Psi}^+_{l,m}(z)\bar{\Psi}_{l,m}(z)\rt)
     -\Psi^+_l(z)\Psi_l(z)\no\\
     &&+\sqrt{k+2r-2n-1}\partial\phi_l(z),\qquad\qquad 1\leq l\leq
     n,\\
  H_{n+i}(z)&=&\hspace{-0.16truecm}\sum_{m=1}^{n}\hspace{-0.16truecm}
     \lt(\Psi^+_{m,i}(z)\Psi_{m,i}(z)
     \hspace{-0.1truecm}-\hspace{-0.1truecm}\bar{\Psi}^+_{m,i}(z)\bar{\Psi}_{m,i}(z)\rt)
     \hspace{-0.1truecm}+\hspace{-0.1truecm}
     \sum_{m=1}^{i-1}\hspace{-0.16truecm}\lt(\g'_{m,i}(z)\b'_{m,i}(z)\hspace{-0.1truecm}
     -\hspace{-0.1truecm}\bar{\g}'_{m,i}(z)\bar{\b}'_{m,i}(z)\rt)\no\\
     &&-\sum_{m=i+1}^r\lt(\g'_{i,m}(z)\b'_{i,m}(z)
     +\bar{\g}'_{i,m}(z)\bar{\b}'_{i,m}(z)\rt)
     -\g'_i(z)\b'_i(z)\no\\
     &&+\sqrt{k+2r-2n-1}\partial\phi_{n+i}(z),\qquad\qquad 1\leq i\leq
     r.\label{Free-SBC-2}
\eea Here normal ordering of free fields is implied.

\end{Theorem}

\vskip0.1in

\noindent {\it Proof}. The proof of the theorem is similar to that
of Theorem 1. \hspace{1.truecm}$\Box$

\vskip0.12in

The free field realization of the currents associated with the
non-simple roots can be obtained from the OPEs of the simple ones.
For the case of $n=0$, our result recovers the free field
realization proposed in \cite{Yan08} for $so(2r+1)$ current
algebra.

\subsubsection{Energy-momentum tensor} \label{EMT-1}

The energy-momentum tensor $T(z)$ associated with the
$osp(2r+1|2n)$ current algebra can be expressed in terms of the
free fields through the Sugawara construction, \bea
  T(z)&=&\frac{1}{2\lt(k+2r-2n-1\rt)}\lt\{-\sum_{m<l}^n\lt(
     E_{\d_m-\d_l}(z)F_{\d_m-\d_l}(z)+F_{\d_m-\d_l}(z)E_{\d_m-\d_l}(z)\rt)\rt.\no\\
  &&-\sum_{m<l}^n\lt(
     E_{\d_m+\d_l}(z)F_{\d_m+\d_l}(z)+F_{\d_m+\d_l}(z)E_{\d_m+\d_l}(z)\rt)\no\\
  &&-\sum_{l=1}^n\lt[\,2\lt(
     E_{2\d_l}(z)F_{2\d_l}(z)+F_{2\d_l}(z)E_{2\d_l}(z)\rt)+H_l(z)H_l(z)\,\rt]\no\\
  &&+\sum_{l=1}^n\lt(E_{\d_l}(z)F_{\d_l}(z)-F_{\d_l}(z)E_{\d_l}(z)\rt)\no\\
  &&+\sum_{l=1}^n\sum_{i=1}^r\lt(
     E_{\d_l-\e_i}(z)F_{\d_l-\e_i}(z)-F_{\d_l-\e_i}(z)E_{\d_l-\e_i}(z)\rt)\no\\
  &&+\sum_{l=1}^n\sum_{i=1}^r\lt(
     E_{\d_l+\e_i}(z)F_{\d_l+\e_i}(z)-F_{\d_l+\e_i}(z)E_{\d_l+\e_i}(z)\rt)\no\\
  &&+\sum_{i<j}^r\lt(E_{\e_i-\e_j}(z)F_{\e_i-\e_j}(z)
     +F_{\e_i-\e_j}(z)E_{\e_i-\e_j}(z)\rt)\no\\
  &&+\sum_{i<j}^r\lt(E_{\e_i+\e_j}(z)F_{\e_i+\e_j}(z)+F_{\e_i+\e_j}(z)E_{\e_i+\e_j}(z)\rt)
      \no\\
  &&+\lt.\sum_{i=1}^r\lt[\lt(E_{\e_i}(z)F_{\e_i}(z)+F_{\e_i}(z)E_{\e_i}(z)\rt)
      +H_{n+i}(z)H_{n+i}(z)\rt]\rt\}.\no
\eea After a tedious calculation, we have

\begin{Proposition}
The energy-momentum tensor $T(z)$ of the $osp(2r+1|2n)$ current
algebra can be expressed  in terms of the  free fields
(\ref{OPE-F-1})-(\ref{OPE-F-2}) and
(\ref{OPE-F-3})-(\ref{OPE-F-4}) as
\bea
  T(z)
  &=&-\sum_{l=1}^n\lt(\frac{1}{2}\partial\phi_l(z)\partial\phi_l(z)-
      \frac{2n+1-2r-2l}{2\sqrt{k+2r-2n-1}}\partial^2\phi_l(z)\rt)\no\\
  &&+\sum_{i=1}^r\lt(\frac{1}{2}\partial\phi_{n+i}(z)\partial\phi_{n+i}(z)-
      \frac{2r-2i+1}{2\sqrt{k+2r-2n-1}}\partial^2\phi_{n+i}(z)\rt)\no\\
  &&+\sum_{m<l}^n\lt(\b_{m,l}(z)\partial\g_{m,l}(z)+
     \bar{\b}_{m,l}(z)\partial\bar{\g}_{m,l}(z)\rt)
     +\sum_{l=1}^n\b_l(z)\partial\g_l(z)\no\\
  &&+\sum_{i<j}^r\lt(\b'_{i,j}(z)\partial\g'_{i,j}(z)+
     \bar{\b}'_{i,j}(z)\partial\bar{\g}'_{i,j}(z)\rt)
     +\sum_{i=1}^r\b'_i(z)\partial\g'_i(z)\no\\
  &&-\sum_{l=1}^n\sum_{i=1}^r\lt(\Psi_{l,i}(z)\partial\Psi^+_{l,i}(z)
     +\bar{\Psi}_{l,i}(z)\partial\bar{\Psi}^+_{l,i}(z)\rt)
     -\sum_{l=1}^n\Psi_l(z)\partial\Psi^+_l(z),\label{Energy-SBC}
\eea where normal ordering of the free fields is implied. $T(z)$
satisfies the OPE of the Virasoro algebra, \bea
   T(z)T(w)=\frac{c/2}{(z-w)^4}+\frac{2T(w)}{(z-w)^2}+\frac{\partial
        T(w)}{(z-w)},
\eea with the central charge $c$ given by
\bea
 c=\frac{(r-n)(2r-2n+1)k}{k+2r-2n-1}\equiv
 \frac{k\times{\rm sdim}\lt(osp(2r+1|2n)\rt)}{k+2r-2n-1}.\label{center-charge-SBC}
\eea
\end{Proposition}

\vskip0.16in

\noindent Note that when $n=r$, i.e. for the $osp(2n+1|2n)$ case,
the central charge (\ref{center-charge-SBC}) vanishes. It can be
easily checked that   the $osp(2r+1|2n)$ currents
(\ref{Free-SBC-1})-(\ref{Free-SBC-2}) are primary fields with
conformal dimensional one, namely, \bea
  T(z)E_{i}(w)&=&\frac{E_{i}(w)}{(z-w)^2}+\frac{\partial
    E_{i}(w)}{(z-w)},\,\,1\leq i\leq n+r,\no\\
  T(z)F_{i}(w)&=&\frac{F_{i}(w)}{(z-w)^2}+\frac{\partial
    F_{i}(w)}{(z-w)},\,\,1\leq i\leq n+r,\no\\
  T(z)H_{i}(w)&=&\frac{H_{i}(w)}{(z-w)^2}+\frac{\partial
    H_{i}(w)}{(z-w)},\,\,1\leq i\leq n+r.\no
\eea

\subsubsection{Screening currents}
Similarly to the $osp(2r|2n)$ case, the free field realization of
the screening currents can be constructed from certain
differential operators defined by the relation, \bea
 \rho^{(d)}\lt(s_{\a}\rt)\,
 \langle\L;x,\bar{x};y,\bar{y};\theta,\bar{\theta}|
 \equiv\langle\L|\,E_{\a}\,
 G_{+}(x,\bar{x};y,\bar{y},\theta,\bar{\theta}),\qquad
 {\rm for}\,\,\a\in\Delta_+,
 \label{Def-SBC-2}
\eea where $\langle\L|$ is given by
(\ref{highestweight-SBC-1})-(\ref{highestweight-SBC-2}) and
$G_{+}(x,\bar{x};y,\bar{y},\theta,\bar{\theta})$ is given by
(\ref{Coordinate-SBC-1})-(\ref{Coordinate-SBC-2}). The operators
$\rho^{(d)}\lt(s_{\a}\rt)$ ($\a\in\Delta_+$) give a differential
operator realization of the subalgebra $osp(2r+1|2n)_+$. After
some algebraic manipulations, we obtain the following explicit
expressions for $s_i\equiv \rho^{(d)}\lt(s_{\a_i}\rt)$: \bea
   s_{l}&=&\sum_{m=l+2}^n\hspace{-0.08truecm}\lt(
           -\bar{x}_{l+1,m}x_{l+1,m}\partial_{\bar{x}_{l,l+1}}
           \hspace{-0.08truecm}+\hspace{-0.08truecm}
           \bar{x}_{l+1,m}\partial_{\bar{x}_{l,m}}
           \hspace{-0.08truecm}+\hspace{-0.08truecm}
           2x_{l+1,m}\bar{x}_{l,m}\partial_{x_l}
           \hspace{-0.08truecm}+\hspace{-0.08truecm}
           x_{l+1,m}\partial_{x_{l,m}}\rt)\no\\
       &&+\sum_{m=1}^r\hspace{-0.08truecm}\lt(
          -\bar{\theta}_{l+1,m}\theta_{l+1,m}\partial_{\bar{x}_{l,l+1}}
           \hspace{-0.08truecm}+\hspace{-0.08truecm}
           \bar{\theta}_{l+1,m}\partial_{\bar{\theta}_{l,m}}
           \hspace{-0.08truecm}-\hspace{-0.08truecm}
           2\theta_{l+1,m}\bar{\theta}_{l,m}\partial_{x_l}
           \hspace{-0.08truecm}+\hspace{-0.08truecm}
           \theta_{l+1,m}\partial_{\theta_{l,m}}\rt)\no\\
      &&+x_{l+1}\partial_{\bar{x}_{l,l+1}}
           \hspace{-0.08truecm}+\hspace{-0.08truecm}
           \theta_{l+1}\partial_{\theta_l}\hspace{-0.08truecm}-\hspace{-0.08truecm}
           \theta_{l+1}\theta_{l}\partial_{x_l}
           \hspace{-0.08truecm}+2\bar{x}_{l,l+1}\partial_{x_l}
           \hspace{-0.08truecm}+\hspace{-0.08truecm}\partial_{x_{l,l+1}},
           \quad 1\leq l\leq n-1,\label{Scr-P-SBC-1}\\
   s_n&=&\sum_{m=2}^r\lt(
           \bar{y}_{1,m}\partial_{\bar{\theta}_{n,m}}
           -\bar{y}_{1,m}y_{1,m}\partial_{\bar{\theta}_{n,1}}
           +y_{1,m}\partial_{\theta_{n,m}}
           -2y_{1,m}\bar{\theta}_{n,m}\partial_{x_n}\rt)\no\\
      &&+y_1\theta_n\partial_{x_n}-y_1\partial_{\theta_n}
           -\frac{y_1^2}{2}\partial_{\bar{\theta}_{n,1}}
           -2\bar{\theta}_{n,1}\partial_{x_n}+\partial_{\theta_{n,1}},\\
   s_{n+i}&=& \sum_{m=i+2}^r\lt(\bar{y}_{i+1,m}\partial_{\bar{y}_{i,m}}
            -\bar{y}_{i+1,m}y_{i+1,m}\partial_{\bar{y}_{i,i+1}}
            +y_{i+1,m}\partial_{y_{i,m}}\rt)\no\\
      &&+y_{i+1}\partial_{y_i}-\frac{y_{i+1}^2}{2}\partial_{\bar{y}_{i,i+1}}
            +\partial_{y_{i,i+1}},\qquad
            1\leq i\leq r-1,\\
   s_{n+r}&=&\partial_{y_{r}}.\label{Scr-P-SBC-2}
\eea Then we have
\begin{Proposition}
The free field realization of the screening currents $S_i(z)$ of
the $osp(2r+1|2n)$ current algebra corresponding to the above
differential operators $s_i$ is given by
 \bea
 S_{l}(z)&=&\lt\{\sum_{m=l+2}^{n}\lt(
        -\bar{\g}_{l+1,m}(z)\g_{l+1,m}(z)\bar{\b}_{l,l+1}(z)
        +\bar{\g}_{l+1,m}(z)\bar{\b}_{l,m}(z)\rt)\rt.\no\\
    &&\quad+\sum_{m=l+2}^{n}\lt(2\g_{l+1,m}(z)\bar{\g}_{l,m}(z)\b_l(z)
        +\g_{l+1,m}(z)\b_{l,m}(z)\rt)+\g_{l+1}(z)\bar{\b}_{l,l+1}(z)\no\\
    &&\quad +2\bar{\g}_{l,l+1}(z)\b_l(z)-\sum_{m=1}^r\lt(
        \bar{\Psi}^+_{l+1,m}(z)\Psi^+_{l+1,m}(z)\bar{\b}_{l,l+1}(z)
        -\bar{\Psi}^+_{l+1,m}(z)\bar{\Psi}_{l,m}(z)\rt)\no\\
    &&\quad+\Psi^+_{l+1}(z)\Psi_l(z)-\Psi^+_{l+1}(z)\Psi^+_{l}(z)\b_l(z)
        +\b_{l,l+1}(z)\no\\
    &&\quad-\lt.\sum_{m=1}^r\lt(
        2\Psi^+_{l+1,m}(z)\bar{\Psi}^+_{l,m}(z)\b_{l}(z)
        -\Psi^+_{l+1,m}(z)\Psi_{l,m}(z)\rt)\rt\}
        e^{\frac{\a_{l}\cdot\vec{\phi}(z)}{\sqrt{k+2r-2n-1}}},\no\\
    &&\qquad\qquad 1\leq l\leq n-1,\label{Screen-SBC-1}\\[6pt]
 S_{n}(z)&=&\lt\{\sum_{m=2}^r\lt(\bar{\g}'_{1,m}(z)\bar{\Psi}_{n,m}(z)
        -\bar{\g}'_{1,m}(z)\g'_{1,m}(z)\bar{\Psi}_{n,1}(z)
        -2\g'_{1,m}(z)\bar{\Psi}^+_{n,m}(z)\b_n(z)\rt)\rt.\no\\
    &&\quad+\g'_1(z)\Psi^+_n(z)\b_n(z)-\g'_1(z)\Psi_n(z)
        -\frac{1}{2}\g'_1(z)\g'_1(z)\bar{\Psi}_{n,1}(z)\no\\
    &&\quad\lt.+\sum_{m=2}^r\g'_{1,m}(z)\Psi_{n,m}(z)
        -2\bar{\Psi}^+_{n,1}(z)\b_n(z)+\Psi_{n,1}(z)\rt\}
        e^{\frac{\a_{_n}\cdot\vec{\phi}(z)}{\sqrt{k+2r-2n-1}}},\\
 S_{n+i}(z)&=&\lt\{\sum_{m=i+2}^r\lt(\bar{\g}'_{i+1,m}(z)\bar{\b}'_{i,m}(z)
        -\bar{\g}'_{i+1,m}(z)\g'_{i+1,m}(z)\bar{\b}'_{i,i+1}(z)
        \rt)\rt.\no\\
    &&\quad
        +\g'_{i+1}(z)\b'_i(z)-\frac{1}{2}\g'_{i+1}(z)\g'_{i+1}(z)\bar{\b}'_{i,i+1}(z)\no\\
    &&\quad+\lt.\sum_{m=i+2}^r\g'_{i+1,m}(z)\b'_{i,m}(z)
        +\b'_{i,i+1}(z)\rt\}e^{\frac{\a_{n+i}\cdot\vec{\phi}(z)}{\sqrt{k+2r-2n-1}}},
        \quad 1\leq i\leq r-1,\\
 S_{n+r}(z)&=&\b'_{r}(z)e^{\frac{\a_{n+r}\cdot\vec{\phi}(z)}{\sqrt{k+2r-2n-1}}},
 \label{Screen-SBC-2}
\eea where normal ordering of free fields is implied and
$\vec{\phi}(z)$ is given by (\ref{Defin-Phi}).
\end{Proposition}

From direct calculation we find that the screening currents
satisfy the required OPEs with the energy-momentum tensor
(\ref{Energy-SBC}) and the $osp(2r+1|2n)$ currents
(\ref{Free-SBC-1})-(\ref{Free-SBC-2}), namely, \bea
  && T(z)S_i(w)=\frac{S_i(w)}{(z-w)^2}+\frac{\partial
       S_i(w)}{(z-w)}=\partial_w\lt\{\frac{S_i(w)}{(z-w)}\rt\},
       \,\,i=1,\ldots,n+r,\\
 &&E_{i}(z)S_j(w)=0,\qquad i,j=1\ldots,n+r,\\
 &&H_i(z)S_j(w)=0,\qquad i,j=1\ldots,n+r,\\
 &&F_i(z)S_j(w)=(-1)^{[[i]]+[F_i]}\d_{ij}\,
\partial_{w}\lt\{\frac{\lt(k+2r-2n-1\rt) \,
e^{\frac{\a_i\cdot\vec{\phi}(w)}{\sqrt{k+2r-2n-1}}}}{(z-w)}\rt\},\no\\
&&\qquad\qquad i,j=1,\ldots,n+r, \eea where $[[i]]$ is given by
(\ref{Def-[]}).

%%%%%%%%%%%%%%%%%%%%%%%%%%%%%%%%%%%%%%%%%%%%%%%%%%%%%%%%%%%%%%%
%                                                             %
%  6. Discussions                                             %
%                                                             %
%%%%%%%%%%%%%%%%%%%%%%%%%%%%%%%%%%%%%%%%%%%%%%%%%%%%%%%%%%%%%%%

\section{Discussions}
\label{Con} \setcounter{equation}{0}

Based on the particular orderings (\ref{order-SD}) and
(\ref{order-SBC}) for the positive roots of the finite dimensional
basic Lie superalgebras, we have constructed the explicit
differential operator realizations
(\ref{Diff-SD-1})-(\ref{Diff-SD-2}) and
(\ref{Diff-SBC-1})-(\ref{Diff-SBC-2}) for the $osp(2r|2n)$ and
$osp(2r+1|2n)$ superalgebras and explicit free field
representations  (\ref{Free-SD-1})-(\ref{Free-SD-2}) and
(\ref{Free-SBC-1})-(\ref{Free-SBC-2}) of their corresponding
current superalgebras. The corresponding energy-momentum tensors
are given in terms of the free fields by (\ref{Energy-SD}) and
(\ref{Energy-SBC}) respectively. We have also found the free field
representations (\ref{Screen-SD-1})-(\ref{Screen-SD-2}) and
(\ref{Screen-SBC-1})-(\ref{Screen-SBC-2}) of the associated
screening currents of the first kind.

These free field realizations of the $osp(2r|2n)$ and
$osp(2r+1|2n)$ current algebras give rise to the Fock
representations of the current algebras. They provide explicit
realizations of the vertex operator construction
\cite{Lep84,Pri02} of representations for affine superalgebras
$osp(2r|2n)_{k}$ and $osp(2r+1|2n)_{k}$. These representations are
in general not irreducible. To obtain irreducible representations,
one needs the associated screening charges, which are the
integrals of the corresponding screening currents
(\ref{Screen-SD-1})-(\ref{Screen-SD-2}) and
(\ref{Screen-SBC-1})-(\ref{Screen-SBC-2}),  and one then performs
the cohomology analysis as in \cite{Fat86,Fei90,Bou90,Ber89}.

An important open problem is to construct the free field
representations of the primary fields for the current
superalgebras studied in this paper. It is well-known that there
exist two types of representations for the underlying
finite-dimensional superalgebras: typical and atypical
representations. Atypical representations have no counterpart in
the bosonic algebra setting and our understanding of such
representations is still very much incomplete. Although the
construction of the primary fields associated with typical
representations is similar to the bosonic algebra cases, it is a
highly nontrivial task to construct the primary fields associated
with atypical representations even for the relatively simple
$gl(2|2)$ current algebra \cite{Zha05}.

%%%%%%%%%%%%%%%%%%%%%%%%%%%%%%%%%%%%%%%%%%%%%%%%%%%%%%%%%%%%%%%
%                                                             %
%  Acknowledgments                                            %
%                                                             %
%%%%%%%%%%%%%%%%%%%%%%%%%%%%%%%%%%%%%%%%%%%%%%%%%%%%%%%%%%%%%%%
\section*{Acknowledgements}
The financial support from  the Australian Research Council is
gratefully acknowledged. WLY has also been partially supported by
the New Staff Research Grant of The University of Queensland.

%%%%%%%%%%%%%%%%%%%%%%%%%%%%%%%%%%%%%%%%%%%%%%%%%%%%%%%%%%%%%%%%%%%%%%%%%%
%                                                                        %
%  Appendix A: Defining representation $osp(2r|2n)$                      %
%                                                                        %
%%%%%%%%%%%%%%%%%%%%%%%%%%%%%%%%%%%%%%%%%%%%%%%%%%%%%%%%%%%%%%%%%%%%%%%%%%

\section*{Appendix A: Defining representation of $osp(m|2n)$}
\setcounter{equation}{0}
\renewcommand{\theequation}{A.\arabic{equation}}
Let $V$ be a $\Zb_2$-graded $(m+2n)$-dimensional vector space with
the orthonornal basis $\{|i\rangle, i=1,\ldots, m+2n\}$. The
$\Zb_2$-grading is chosen as: $[1]=\cdots=[m]=0,\,
[m+1]=\cdots=[m+2n]=1$. For any $(m+2n)\times (m+2n)$ matrix $A$,
one can define the supertrace,
 \bea
 str\lt(A\rt)=\sum_{l=1}^{m+2n}(-1)^{[l]}A_{ll}
 =\sum_{l=1}^{m}A_{ll}-\sum_{l=m+1}^{m+2n}A_{ll}.\label{supertrace}
\eea

Let $e_{ij}$, $i,j=1,\ldots,n$, be an $n\times n$ matrix with
entry $1$ at the $i$th row and the $j$th column and zero
elsewhere. Let $e_i$, $i=1,\ldots,n$, be an $n$-dimensional row
vector with the $i$th component being $1$ and all others being
zero, and $e_i^T$ be the transpose of $e_i$, namely, \bea
 e_i=(0,\ldots,0,1,0,\ldots,0),\qquad e_i^T=\lt(\begin{array}{l}0\\
   \vdots\\0\\1\\0\\\vdots\\0\end{array}\rt).\no
\eea Similarly, one can introduce  $r\times r$ matrices
$\bar{e}_{ij}$, $i,j=1,\ldots,r$, and $r$-dimensional row vectors
$\bar{e}_i$. With help of these matrices
$\{e_{ij}|\,i,j=1,\ldots,n\}$ and
$\{\bar{e}_{ij}|\,i,j=1,\ldots,r\}$ and row vectors
$\{e_i|\,i=1,\ldots,n\}$ and $\{\bar{e}_i|\,i=1,\ldots,r\}$, one
can realize the defining representations of $osp(2r|2n)$ and
$osp(2r+1|2n)$ as follows.

%%%%%%%%%%%%%%%%%%%%%%%%%%%%%%%%%%%%%%%%%%%%%%%%%%%%%%%%%%%%%%%%%%
\subsection*{A1. The $osp(2r|2n)$ case}

Let $m=2r$, i.e. dim(V)$=2(r+n)$. The defining representation of
$osp(2r|2n)$ on V, denoted by $\rho_{0}$, is given by the
following $2(r+n)\times 2(r+n)$ matrices, \bea
\hspace{-1.26truecm}
 \rho_{0}\lt(E_{\d_m-\d_l}\rt)
   \hspace{-0.32truecm}&=&\hspace{-0.32truecm}
   \left(\begin{array}{c|c}{\,\,\,\,\,\,}&{\,\,\,\,\,\,}\\[6pt]
   \hline &{\begin{array}{cc}e_{ml}&\\
   &-e_{lm}\end{array}}\end{array}\right),\,\,
 \rho_{0}\lt(F_{\d_m-\d_l}\rt)=
   \left(\begin{array}{c|c}{\,\,\,\,\,\,}&{\,\,\,\,\,\,}\\[6pt]
   \hline &{\begin{array}{cc}e_{lm}&\\
   &-e_{ml}\end{array}}\end{array}\right),\,\, m<l,\label{F-R-SD-1}\\[8pt]
\hspace{-1.26truecm}
 \rho_{0}\lt(E_{2\d_l}\rt)
 \hspace{-0.32truecm}&=&\hspace{-0.32truecm}
   \left(\begin{array}{c|c}{\,\,\,\,\,\,}&{\,\,\,\,\,\,}\\[6pt]
   \hline &{\begin{array}{cc}0&e_{ll}\\
   0&0\end{array}}\end{array}\right),\,\,
 \rho_{0}\lt(F_{2\d_l}\rt)=
   \left(\begin{array}{c|c}{\,\,\,\,\,\,}&{\,\,\,\,\,\,}\\[6pt]
   \hline &{\begin{array}{cc}0&0\\
   e_{ll}&0\end{array}}\end{array}\right),\\[8pt]
\hspace{-1.26truecm}
 \rho_{0}\lt(E_{\d_m+\d_l}\rt)
   \hspace{-0.32truecm}&=&\hspace{-0.32truecm}
   \left(\begin{array}{c|c}{\,\,\,\,\,\,}&{\,\,\,\,\,\,}\\[6pt]
   \hline &{\begin{array}{cc}0&e_{ml}+e_{lm}\\
   0&0\end{array}}\end{array}\right),\,\,
 \rho_{0}\lt(F_{\d_m+\d_l}\rt)=
   \left(\begin{array}{c|c}{\,\,\,\,\,\,}&{\,\,\,\,\,\,}\\[6pt]
   \hline &{\begin{array}{cc}0&0\\
   e_{ml}+e_{lm}&0\end{array}}\end{array}\right),\,\, m<l,\\[8pt]
\hspace{-1.26truecm}
 \rho_{0}\lt(E_{\d_l-\e_i}\rt)
   \hspace{-0.32truecm}&=&\hspace{-0.32truecm}
   \left(\begin{array}{c|c}&{\begin{array}{cc}0&0\\
   0&\bar{e}^T_ie_{l}\end{array}}\\[6pt]
   \hline {\begin{array}{cc}e^T_l\bar{e}_{i}&0\\0&0\end{array}}&\\
   \end{array}\right),\,\,
 \rho_{0}\lt(F_{\d_l-\e_i}\rt)=
   \left(\begin{array}{c|c}&{\begin{array}{cc}\bar{e}^T_ie_{l}&0\\
   0&0\end{array}}\\[6pt]
   \hline {\begin{array}{cc}0&0\\0&-e^T_l\bar{e}_{i}\end{array}}&\\
   \end{array}\right),\\[8pt]
\hspace{-1.26truecm}
 \rho_{0}\lt(E_{\d_l+\e_i}\rt)
   \hspace{-0.32truecm}&=&\hspace{-0.32truecm}
   \left(\begin{array}{c|c}&{\begin{array}{cc}0&\bar{e}^T_ie_{l}\\
   0&0\end{array}}\\[6pt]
   \hline {\begin{array}{cc}0&e^T_l\bar{e}_{i}\\0&0\end{array}}&\\
   \end{array}\right),\,\,
 \rho_{0}\lt(F_{\d_l+\e_i}\rt)=
   \left(\begin{array}{c|c}&{\begin{array}{cc}0&0\\
   \bar{e}^T_ie_{l}&0\end{array}}\\[6pt]
   \hline {\begin{array}{cc}0&0\\-e^T_l\bar{e}_{i}&0\end{array}}&\\
   \end{array}\right),\\[8pt]
\hspace{-1.26truecm}
 \rho_{0}\lt(E_{\e_i-\e_j}\rt)
   \hspace{-0.32truecm}&=&\hspace{-0.32truecm}
   \left(\begin{array}{c|c}{\begin{array}{cc}\bar{e}_{ij}&\\
   &-\bar{e}_{ji}\end{array}}&\\
   [8pt]\hline {\,\,\,\,\,\,}&{\,\,\,\,\,\,}\end{array}\right),\,\,
 \rho_{0}\lt(F_{\e_i-\e_j}\rt)=
   \left(\begin{array}{c|c}{\begin{array}{cc}\bar{e}_{ji}&\\
   &-\bar{e}_{ij}\end{array}}&\\
   [8pt]\hline {\,\,\,\,\,\,}&{\,\,\,\,\,\,}\end{array}\right),\qquad\quad i<j,\\[8pt]
\hspace{-1.26truecm}
 \rho_{0}\lt(E_{\e_i+\e_j}\rt)
   \hspace{-0.32truecm}&=&\hspace{-0.32truecm}
   \left(\begin{array}{c|c}{\begin{array}{cc}0&\bar{e}_{ij}-\bar{e}_{ji}\\
   0&0\end{array}}&\\
   [8pt]\hline {\,\,\,\,\,\,}&{\,\,\,\,\,\,}\end{array}\right),\,\,
 \rho_{0}\lt(F_{\e_i+\e_j}\rt)=
   \left(\begin{array}{c|c}{\begin{array}{cc}0&0\\-\bar{e}_{ij}+\bar{e}_{ji}&0\end{array}}&\\
   [8pt]\hline {\,\,\,\,\,\,}&{\,\,\,\,\,\,}\end{array}\right),\,\, i<j,\\[8pt]
\hspace{-1.26truecm}
 \rho_{0}\lt(H_{\d_m-\d_l}\rt)
   \hspace{-0.32truecm}&=&\hspace{-0.32truecm}
   \left(\begin{array}{c|c}{\,\,\,\,\,\,}&{\,\,\,\,\,\,}\\[6pt]
   \hline &{\begin{array}{cc}e_{mm}-e_{ll}&\\
   &e_{ll}-e_{mm}\end{array}}\end{array}\right),\qquad\qquad m<l,\\[8pt]
\hspace{-1.26truecm}
 \rho_{0}\lt(H_{\d_m+\d_l}\rt)
   \hspace{-0.32truecm}&=&\hspace{-0.32truecm}
   \left(\begin{array}{c|c}{\,\,\,\,\,\,}&{\,\,\,\,\,\,}\\[6pt]
   \hline &{\begin{array}{cc}e_{mm}+e_{ll}&\\
   &-e_{mm}-e_{ll}\end{array}}\end{array}\right),\qquad\qquad m<l,\\[8pt]
\hspace{-1.26truecm}
 \rho_{0}\lt(H_{2\d_l}\rt)
   \hspace{-0.32truecm}&=&\hspace{-0.32truecm}
   \left(\begin{array}{c|c}{\,\,\,\,\,\,}&{\,\,\,\,\,\,}\\[6pt]
   \hline &{\begin{array}{cc}e_{ll}&\\
   &-e_{ll}\end{array}}\end{array}\right),\\[8pt]
\hspace{-1.26truecm}
 \rho_{0}\lt(H_{\d_l-\e_i}\rt)
   \hspace{-0.32truecm}&=&\hspace{-0.32truecm}
   \left(\begin{array}{c|c}{\begin{array}{cc}\bar{e}_{ii}&\\&-\bar{e}_{ii}\end{array}}&\\[6pt]
   \hline &{\begin{array}{cc}e_{ll}&\\
   &-e_{ll}\end{array}}\end{array}\right),\,\,
 \rho_{0}\lt(H_{\d_l+\e_i}\rt)\hspace{-0.1truecm}=\hspace{-0.1truecm}
   \left(\begin{array}{c|c}{\begin{array}{cc}-\bar{e}_{ii}&\\&\bar{e}_{ii}\end{array}}&\\[6pt]
   \hline &{\begin{array}{cc}e_{ll}&\\
   &-e_{ll}\end{array}}\end{array}\right),\\[8pt]
\hspace{-1.26truecm}
 \rho_{0}\lt(H_{\e_i-\e_j}\rt)
   \hspace{-0.32truecm}&=&\hspace{-0.32truecm}
   \left(\begin{array}{c|c}{\begin{array}{cc}\bar{e}_{ii}-\bar{e}_{jj}&\\
   &\bar{e}_{jj}-\bar{e}_{ii}\end{array}}&\\[6pt]
   \hline {\,\,\,\,}&{\,\,\,\,}\end{array}\right),\qquad\qquad i<j,\\[8pt]
\hspace{-1.26truecm}
 \rho_{0}\lt(H_{\e_i+\e_j}\rt)
   \hspace{-0.32truecm}&=&\hspace{-0.32truecm}
   \left(\begin{array}{c|c}{\begin{array}{cc}\bar{e}_{ii}+\bar{e}_{jj}&\\
   &-\bar{e}_{ii}-\bar{e}_{jj}\end{array}}&\\[6pt]
   \hline {\,\,\,\,}&{\,\,\,\,}\end{array}\right),\qquad\qquad
   i<j. \label{F-R-SD-2}
\eea  Then we introduce $r+n$ linear-independent generators $H_i$
$(i=1,\ldots r+n)$, \bea
 H_l&=&H_{2\d_l},\qquad 1\leq l\leq n,\label{SD-H-1}\\
 H_{n+i}&=&\frac{1}{2}(H_{\e_i-\e_j}+H_{\e_i+\e_j}),\qquad i=1,\ldots,
   r-1,\,{\rm and}\,\,i<j,\\
 H_{n+r}&=&\frac{1}{2}\lt(H_{\e_i+\e_{r}}-H_{\e_i-\e_{r}}\rt), \qquad i\leq r-1.\label{SD-H-2}
\eea Actually, the above generators $\{H_i\}$ span the Cartan
subalgebra of $osp(2r|2n)$. In the defining representation, these
generators can be realized by \bea
 \rho_{0}\lt(H_l\rt)&=&
   \left(\begin{array}{c|c}{\,\,\,\,\,\,}&{\,\,\,\,\,\,}\\[6pt]
   \hline &{\begin{array}{cc}e_{ll}&\\
   &-e_{ll}\end{array}}\end{array}\right),\,\,
   l=1,\ldots,n,\\[8pt]
 \rho_{0}\lt(H_{n+i}\rt)&=&
   \left(\begin{array}{c|c}{\begin{array}{cc}\bar{e}_{ii}&\\
   &-\bar{e}_{ii}\end{array}}\\[6pt]
   \hline {\,\,\,\,} &{\,\,\,\,}\end{array}\right),\,\,
   i=1,\ldots,r.
\eea

The corresponding nondegenerate invariant bilinear supersymmetric
form of $osp(2r|2n)$ is given by
\bea
 (x,y)=\frac{1}{2}str\lt(\rho_0(x)\rho_0(y)\rt),\qquad
  \forall x,y\in osp(2r|2n).\label{Bilinear-SD}
\eea

%%%%%%%%%%%%%%%%%%%%%%%%%%%%%%%%%%%%%%%%%%%%%%%%%%%%%%%%%%%%%%%%%%
\subsection*{A2. The $osp(2r+1|2n)$ case}

Let $m=2r+1$, i.e. dim(V)$=2(r+n)+1$.The defining representation
of $osp(2r+1|2n)$ on V, denoted by $\rho_{0}$, is given by the
following $(2(r+n)+1)\times (2(r+n)+1)$ matrices, \bea
\hspace{-1.26truecm}
 \rho_{0}\lt(E_{\d_m-\d_l}\rt)
   \hspace{-0.32truecm}&=&\hspace{-0.32truecm}
   \left(\begin{array}{c|c}{\,\,\,\,\,\,}&{\,\,\,\,\,\,}\\[6pt]
   \hline &{\begin{array}{cc}e_{ml}&\\
   &-e_{lm}\end{array}}\end{array}\right),\,\,
 \rho_{0}\lt(F_{\d_m-\d_l}\rt)=
   \left(\begin{array}{c|c}{\,\,\,\,\,\,}&{\,\,\,\,\,\,}\\[6pt]
   \hline &{\begin{array}{cc}e_{lm}&\\
   &-e_{ml}\end{array}}\end{array}\right),\,\, m<l,\label{F-R-SBC-1}\\[8pt]
\hspace{-1.26truecm}
 \rho_{0}\lt(E_{2\d_l}\rt)
 \hspace{-0.32truecm}&=&\hspace{-0.32truecm}
   \left(\begin{array}{c|c}{\,\,\,\,\,\,}&{\,\,\,\,\,\,}\\[6pt]
   \hline &{\begin{array}{cc}0&e_{ll}\\
   0&0\end{array}}\end{array}\right),\,\,
 \rho_{0}\lt(F_{2\d_l}\rt)=
   \left(\begin{array}{c|c}{\,\,\,\,\,\,}&{\,\,\,\,\,\,}\\[6pt]
   \hline &{\begin{array}{cc}0&0\\
   e_{ll}&0\end{array}}\end{array}\right),\\[8pt]
\hspace{-1.26truecm}
 \rho_{0}\lt(E_{\d_m+\d_l}\rt)
   \hspace{-0.32truecm}&=&\hspace{-0.32truecm}
   \left(\begin{array}{c|c}{\,\,\,\,\,\,}&{\,\,\,\,\,\,}\\[6pt]
   \hline &{\begin{array}{cc}0&e_{ml}+e_{lm}\\
   0&0\end{array}}\end{array}\right)\hspace{-0.12truecm},\,\,
 \rho_{0}\lt(F_{\d_m+\d_l}\rt)\hspace{-0.12truecm}=\hspace{-0.12truecm}
   \left(\begin{array}{c|c}{\,\,\,\,\,\,}&{\,\,\,\,\,\,}\\[6pt]
   \hline &{\begin{array}{cc}0&0\\
   e_{ml}+e_{lm}&0\end{array}}\end{array}\right),\,\, m<l,\\[8pt]
\hspace{-1.26truecm}
 \rho_{0}\lt(E_{\d_l-\e_i}\rt)
   \hspace{-0.32truecm}&=&\hspace{-0.32truecm}
   \left(\begin{array}{c|c}&{\begin{array}{cc}0&0\\0&0\\
   0&\bar{e}^T_ie_{l}\end{array}}\\[6pt]
   \hline {\begin{array}{ccc}0&e^T_l\bar{e}_{i}&0\\0&0&0\end{array}}&\\
   \end{array}\right)\hspace{-0.12truecm},\,
 \rho_{0}\hspace{-0.12truecm}\lt(F_{\d_l-\e_i}\rt)
   \hspace{-0.12truecm}=\hspace{-0.12truecm}\hspace{-0.12truecm}
   \left(\begin{array}{c|c}&{\begin{array}{cc}0&0\\\bar{e}^T_ie_{l}&0\\
   0&0\end{array}}\\[6pt]
   \hline {\begin{array}{ccc}0&0&0\\0&0&-e^T_l\bar{e}_{i}\end{array}}&\\
   \end{array}\right)\hspace{-0.12truecm},\\[8pt]
\hspace{-1.26truecm}
 \rho_{0}\lt(E_{\d_l+\e_i}\rt)
   \hspace{-0.32truecm}&=&\hspace{-0.32truecm}
   \left(\begin{array}{c|c}&{\begin{array}{cc}0&0\\0&\bar{e}^T_ie_{l}\\
   0&0\end{array}}\\[6pt]
   \hline {\begin{array}{ccc}0&0&e^T_l\bar{e}_{i}\\0&0&0\end{array}}&\\
   \end{array}\right)\hspace{-0.12truecm},\,
 \rho_{0}\hspace{-0.12truecm}\lt(F_{\d_l+\e_i}\rt)
   \hspace{-0.12truecm}=\hspace{-0.12truecm}\hspace{-0.12truecm}
   \left(\begin{array}{c|c}&{\begin{array}{cc}0&0\\0&0\\
   \bar{e}^T_ie_{l}&0\end{array}}\\[6pt]
   \hline {\begin{array}{ccc}0&0&0\\0&-e^T_l\bar{e}_{i}&0\end{array}}&\\
   \end{array}\right)\hspace{-0.12truecm},\\[8pt]
\hspace{-1.26truecm}
 \rho_{0}\lt(E_{\d_l}\rt)
   \hspace{-0.32truecm}&=&\hspace{-0.32truecm}
   \left(\begin{array}{c|c}&{\begin{array}{cc}0&e_l\\0&0\\
   0&0\end{array}}\\[6pt]
   \hline {\begin{array}{ccc}e^T_l&0&0\\0&0&0\end{array}}&\\
   \end{array}\right),\,\,
 \rho_{0}\lt(F_{\d_l}\rt)=
   \left(\begin{array}{c|c}&{\begin{array}{cc}e_l&0\\0&0\\
   0&0\end{array}}\\[6pt]
   \hline {\begin{array}{ccc}0&0&0\\-e^T_l&0&0\end{array}}&\\
   \end{array}\right),\\[8pt]
\hspace{-1.26truecm}
 \rho_{0}\lt(E_{\e_i-\e_j}\rt)
   \hspace{-0.32truecm}&=&\hspace{-0.32truecm}
   \left(\begin{array}{c|c}{\begin{array}{ccc}0&&\\&\bar{e}_{ij}&\\
   &&-\bar{e}_{ji}\end{array}}&\\
   [8pt]\hline {\,\,\,\,\,\,}&{\,\,\,\,\,\,}\end{array}\right)\hspace{-0.12truecm},\,\,
 \rho_{0}\lt(F_{\e_i-\e_j}\rt)\hspace{-0.12truecm}=\hspace{-0.12truecm}
   \left(\begin{array}{c|c}{\begin{array}{ccc}0&&\\&\bar{e}_{ji}&\\
   &&-\bar{e}_{ij}\end{array}}&\\
   [8pt]\hline {\,\,\,\,\,\,}&{\,\,\,\,\,\,}\end{array}\right),\quad i<j,\\[8pt]
\hspace{-1.26truecm}
 \rho_{0}\lt(E_{\e_i+\e_j}\rt)
   \hspace{-0.32truecm}&=&\hspace{-0.32truecm}
   \left(\begin{array}{c|c}{\begin{array}{ccc}0&0&0\\0&0&\bar{e}_{ij}-\bar{e}_{ji}\\
   0&0&0\end{array}}&\\
   [8pt]\hline {\,\,\,\,\,\,}&{\,\,\,\,\,\,}\end{array}\right)\hspace{-0.12truecm},\,
 \rho_{0}\hspace{-0.12truecm}\lt(F_{\e_i+\e_j}\rt)\hspace{-0.12truecm}
   =\hspace{-0.12truecm}\hspace{-0.12truecm}
   \left(\begin{array}{c|c}{\begin{array}{ccc}0&0&0\\0&0&0\\0&-\bar{e}_{ij}+\bar{e}_{ji}&0\end{array}}&\\
   [8pt]\hline {\,\,\,\,\,\,}&{\,\,\,\,\,\,}\end{array}\right)\hspace{-0.12truecm},\,
   i\hspace{-0.12truecm}<\hspace{-0.12truecm}j,\\[8pt]
\hspace{-1.26truecm}
 \rho_{0}\lt(E_{\e_i}\rt)
   \hspace{-0.32truecm}&=&\hspace{-0.32truecm}
   \left(\begin{array}{c|c}{\begin{array}{ccc}0&0&\bar{e}_i\\-\bar{e}^T_i&0&0\\
   0&0&0\end{array}}&\\
   [8pt]\hline
   {\,\,\,\,\,\,}&{\,\,\,\,\,\,}\end{array}\right),\,\,\,\,
 \rho_{0}\lt(F_{\e_i}\rt)=
   \left(\begin{array}{c|c}{\begin{array}{ccc}0&-\bar{e}_i&0\\0&0&0\\
   \bar{e}^T_i&0&0\end{array}}&\\
   [8pt]\hline {\,\,\,\,\,\,}&{\,\,\,\,\,\,}\end{array}\right),\\[8pt]
\hspace{-1.26truecm}
 \rho_{0}\lt(H_{\d_m-\d_l}\rt)
   \hspace{-0.32truecm}&=&\hspace{-0.32truecm}
   \left(\begin{array}{c|c}{\,\,\,\,\,\,}&{\,\,\,\,\,\,}\\[6pt]
   \hline &{\begin{array}{cc}e_{mm}-e_{ll}&\\
   &e_{ll}-e_{mm}\end{array}}\end{array}\right),\qquad\qquad m<l,\\[8pt]
\hspace{-1.26truecm}
 \rho_{0}\lt(H_{\d_m+\d_l}\rt)
   \hspace{-0.32truecm}&=&\hspace{-0.32truecm}
   \left(\begin{array}{c|c}{\,\,\,\,\,\,}&{\,\,\,\,\,\,}\\[6pt]
   \hline &{\begin{array}{cc}e_{mm}+e_{ll}&\\
   &-e_{mm}-e_{ll}\end{array}}\end{array}\right),\qquad\qquad m<l,\\[8pt]
\hspace{-1.26truecm}
 \rho_{0}\lt(H_{2\d_l}\rt)
   \hspace{-0.32truecm}&=&\hspace{-0.32truecm}
   \left(\begin{array}{c|c}{\,\,\,\,\,\,}&{\,\,\,\,\,\,}\\[6pt]
   \hline &{\begin{array}{cc}e_{ll}&\\
   &-e_{ll}\end{array}}\end{array}\right),\,
   \rho_{0}\lt(H_{\d_l-\e_i}\rt)=
   \left(\begin{array}{c|c}{\begin{array}{ccc}0&&\\&\bar{e}_{ii}&\\&&-\bar{e}_{ii}\end{array}}&\\[6pt]
   \hline &{\begin{array}{cc}e_{ll}&\\
   &-e_{ll}\end{array}}\end{array}\right),\\[8pt]
\hspace{-1.26truecm}
 \rho_{0}\lt(H_{\d_l+\e_i}\rt)\hspace{-0.32truecm}&=&\hspace{-0.32truecm}
   \left(\begin{array}{c|c}{\begin{array}{ccc}0&&\\&-\bar{e}_{ii}&\\&&\bar{e}_{ii}\end{array}}&\\[6pt]
   \hline &{\begin{array}{cc}e_{ll}&\\
   &-e_{ll}\end{array}}\end{array}\right),\,
 \rho_{0}\lt(H_{\d_l}\rt)=
   \left(\begin{array}{c|c}{\,\,\,\,\,\,}&{\,\,\,\,\,\,}\\[6pt]
   \hline &{\begin{array}{cc}e_{ll}&\\
   &-e_{ll}\end{array}}\end{array}\right),\\[8pt]
\hspace{-1.26truecm}
 \rho_{0}\lt(H_{\e_i-\e_j}\rt)
   \hspace{-0.32truecm}&=&\hspace{-0.32truecm}
   \left(\begin{array}{c|c}{\begin{array}{ccc}0&&\\&\bar{e}_{ii}-\bar{e}_{jj}&\\
   &&\bar{e}_{jj}-\bar{e}_{ii}\end{array}}&\\[6pt]
   \hline {\,\,\,\,}&{\,\,\,\,}\end{array}\right),\qquad\qquad i<j,\\[8pt]
\hspace{-1.26truecm}
 \rho_{0}\lt(H_{\e_i+\e_j}\rt)
   \hspace{-0.32truecm}&=&\hspace{-0.32truecm}
   \left(\begin{array}{c|c}{\begin{array}{ccc}0&&\\&\bar{e}_{ii}+\bar{e}_{jj}&\\
   &&-\bar{e}_{ii}-\bar{e}_{jj}\end{array}}&\\[6pt]
   \hline {\,\,\,\,}&{\,\,\,\,}\end{array}\right),\qquad\qquad
   i<j,\\[8pt]
\hspace{-1.26truecm}
 \rho_{0}\lt(H_{\e_i}\rt)
   \hspace{-0.32truecm}&=&\hspace{-0.32truecm}
   \left(\begin{array}{c|c}{\begin{array}{ccc}0&&\\&\bar{e}_{ii}&\\
   &&-\bar{e}_{ii}\end{array}}&\\[6pt]
   \hline {\,\,\,\,}&{\,\,\,\,}\end{array}\right).
   \label{F-R-SBC-2}
\eea
We introduce $r+n$ linear-independent generators $H_i$
$(i=1,\ldots r+n)$, \bea
 H_l&=&H_{2\d_l},\qquad 1\leq l\leq n,\label{SBC-H-1}\\
 H_{n+i}&=&\frac{1}{2}(H_{\e_i-\e_j}+H_{\e_i+\e_j}),\qquad i=1,\ldots,
   r-1,\,{\rm and}\,\,i<j,\\
 H_{n+r}&=&\frac{1}{2}\lt(H_{\e_i+\e_{r}}-H_{\e_i-\e_{r}}\rt),
 \qquad i\leq r-1.\label{SBC-H-2}
\eea Actually, the above generators $\{H_i\}$ span the Cartan
subalgebra of $osp(2r+1|2n)$. In the defining representation,
these generators can be realized by \bea
 \rho_{0}\lt(H_l\rt)&=&
   \left(\begin{array}{c|c}{\,\,\,\,\,\,}&{\,\,\,\,\,\,}\\[6pt]
   \hline &{\begin{array}{cc}e_{ll}&\\
   &-e_{ll}\end{array}}\end{array}\right),\,\,
   l=1,\ldots,n,\\[8pt]
 \rho_{0}\lt(H_{n+i}\rt)&=&
   \left(\begin{array}{c|c}{\begin{array}{ccc}0&&\\&\bar{e}_{ii}&\\
   &&-\bar{e}_{ii}\end{array}}\\[6pt]
   \hline {\,\,\,\,} &{\,\,\,\,}\end{array}\right),\,\,
   i=1,\ldots,r.
\eea

The corresponding nondegenerate invariant bilinear supersymmetric
form of $osp(2r+1|2n)$ is given by \bea
 (x,y)=\frac{1}{2}str\lt(\rho_0(x)\rho_0(y)\rt),\qquad
  \forall x,y\in osp(2r+1|2n).\label{Bilinear-SBC}
\eea

%%%%%%%%%%%%%%%%%%%%%%%%%%%%%%%%%%%%%%%%%%%%%%%%%%%%%%%%%%%%%%%
%                                                             %
%  References                                                 %
%                                                             %
%%%%%%%%%%%%%%%%%%%%%%%%%%%%%%%%%%%%%%%%%%%%%%%%%%%%%%%%%%%%%%%


\begin{thebibliography}{99}
\bibitem{Ber99} N. Berkovits, C. Vafa and E. Witten, {\it JHEP\/}
      {\bf 03} (1999), 018.
\bibitem{Bers99} M. Bershadsky, S. Zhukov and A. Vaintrob,
      {\it Nucl. Phys.\/} {\bf B 559} (1999), 205.
\bibitem{Roz92} L. Rozansky and H. Saleur, {\it Nucl. Phys.\/}
      {\bf B 376} (1992), 461.
\bibitem{Gur93} V. Gurarie, {\it Nucl. Phys.\/} {\bf B 410}
      (1993), 535.
\bibitem{Flo03} M. Flohr, {\it Int. J. Mod. Phys. \/} {\bf A 18}
      (2003), 4497.
\bibitem{Gab03} M. Gaberdiel, {\it Int. J. Mod. Phys.\/}
      {\bf A 18} (2003), 4593.
\bibitem{Efe83} K. Efetov, {\it Adv. Phys.\/} {\bf 32}
      (1983), 53.
\bibitem{Ber95}D. Bernard, {\tt hep-th/9509137}.
\bibitem{Mud96} C. Mudry, C. Chamon and X.\,-G. Wen,
      {\it Nucl. Phys.\/} {\bf B 466} (1996), 383.
\bibitem{Maa97} Z. Maassarani and D. Serban, {\it Nucl. Phys.\/}
      {\bf B 489} (1997), 603.
\bibitem{Bas00} Z.\,S. Bassi and A. LeClair, {\it Nucl. Phys.\/}
      {\bf B 578} (2000), 577.
\bibitem{Gur00} S. Guruswamy, A. LeClair and A.\,W.\,W. Ludwig,
      {\it Nucl. Phys.\/} {\bf B 583} (2000), 475.
\bibitem{Lud00} A.\,W.\,W. Ludwig, {\tt cond-mat/0012189}.
\bibitem{Bha01} M.\,J. Bhaseen, J.\,-S. Caux, I.\,I. Kogan and
     A.\,M. Tsveilk, {\it Nucl. Phys.\/} {\bf B 618} (2001), 465.
\bibitem{Met98} R.\,R. Metsaev and A.\,A. Tseytlin, {\it Nucl.
Phys.\/} {\bf B 533} (1998), 109.
\bibitem{Ber00} N. Berkovits, M. Bershadsky, T. Hauer, S. Zhukov
and B. Zwiebach, {\it Nucl. Phys.\/} {\bf B 567} (2000), 61.
\bibitem{Kag06} D. Kagan and C.\,A.\,S. Young,
{\it Nucl. Phys.\/} {\bf B  745} (2006), 109.
\bibitem{Bab07} A. Babichenko, {\it Phys. Lett\/} {\bf B 648}
(2007 ), 254.
\bibitem{Kac90} V. Kac, {\it Infinite-dimensional Lie algebras\/},
     Cambridge University Press, 1990.
\bibitem{Sch06} V. Schomerus and H. Saleur, {\it Nucl. Phys. \/}
      {\bf B 734} (2006), 221; {\it Nucl. Phys.\/} {\bf B 775}
      (2007), 312.
\bibitem{Que07} T. Quella and V. Schomerus, {\it JHEP} {\bf 09}
(2007), 085.
\bibitem{Kac77} V. Kac, {\it Adv. Math.\/} {\bf 26} (1977), 8.
\bibitem{Bel84} A.\,A. Belavin, A.\,M. Polyakov and
     A.\,B. Zamolodchikov, {\it Nucl. Phys.\/} {\bf B 241} (1984), 333.
\bibitem{Fra97} P. Di Francesco, P. Mathieu and D. Senehal, {\it
     Conformal Field Theory\/}, Springer Press, Berlin, 1997.
\bibitem{Sem03} A.\,M. Semikhatov, A. Taormina and I. Yu Timpunin,
      {\it Commun. Math. Phys.\/} {\bf 255} (2005), 469.
\bibitem{Wak86} M. Wakimoto, {\it Commun. Math. Phys.\/}
     {\bf 104} (1986), 605.
\bibitem{Dos84} VI.\,S. Dotsenko and V.\,A. Fateev,
     {\it Nucl. Phys.\/} {\bf B 240} (1984), 312;
     {\it Nucl. Phys.\/} {\bf B 251} (1985), 3691.
\bibitem{Fat86} V.\,A. Fateev and A.\,B. Zamolodchikov,
     {\it Sov. J. Nucl. Phys.\/} {\bf 43} (1986), 657.
\bibitem{God86} P. Goddard, A. kent and D. Olive,
     {\it Phys. Lett.\/} {\bf B 321} (1985), 88;
     {\it Commun. Math. Phys.\/} {\bf 103} (1986), 105.
\bibitem{Ber90} D. Bernard and G. Felder,
     {\it Commun. Math. Phys.\/} {\bf 127} (1990), 145.
\bibitem{Fur93} P. Furlan, A.\,C. Ganchev, R. Paunov and
     V.\,B. Petkova, {\it Nucl. Phys.\/} {\bf B 394}
     (1993), 665.
\bibitem{And95} O. Andreev, {\it Phys. Lett.\/}
     {\bf B 363} (1995), 166.
\bibitem{Fei90} B. Feigin and E. Frenkel,
     {\it Commun. Math. Phys.\/} {\bf 128} (1990), 161; {\it
     Lett. Math. Phys.\/} {\bf 19} (1990), 307.
\bibitem{Ber89} M. Bershadsky and H. Ooguri, {\it Phys. Lett.\/}
     {\bf B 229} (1989), 374;
     {\it Commun. Math. Phys.\/} {\bf 126} (1989), 49.
\bibitem{Bou90} P. Bouwknegt, J. McCarthy and K. Pilch,
     {\it Prog. Theor. Phys. Suppl.\/} {\bf  102} (1990), 67.
\bibitem{Ger90} A. Gerasimov, A. Morozov, M. Olshanetsky,
     A. Marshakov and S. Shatashvili, {\it Int. J. Mod. Phys.\/}
     {\bf A 5} (1990), 2495.
\bibitem{Ito91} K. Ito and S. Komata, {\it Mod. Phys. Lett.\/}
     {\bf A 6} (1991), 581.
%\bibitem{Fre94} E. Frenkel, {\tt hep-th/9408109}.
\bibitem{Boe97} J. de Boer and L. Feher, {\it Mod. Phys. Lett.\/}
      {\bf A 11} (1996), 1999; {\it Commun. Math. Phys.\/} {\bf 189} (1997), 759.
%\bibitem{Ras97} J.\,L. Peterson, J. Rasmussen and M. Yu,
%     {\it Nucl. Phys.\/} {\bf B 502} (1997), 649.
\bibitem{Ras98} J. Rasmussen, {\it Nucl. Phys.\/} {\bf B 510}
     (1998), 688.
\bibitem{Fre06} E. Frenkel, {\it QFT and Geometric Lanlands Program,
      Langlands Correspondence For Loop Groups An Introduction}.
\bibitem{Bow96} P. Bowcock, R-L.\,K. Koktava and A. Taormina,
     {\it Phys. Lett.\/} {\bf B 388} (1996), 303.
\bibitem{Din03} X.\,-M. Ding, M. Gould and Y.\,-Z. Zhang,
     {\it Phys. Lett.\/} {\bf A 318} (2003), 354.
\bibitem{Din03-1} X.\,-M. Ding, M.\,D. Gould, C.\,J. Mewton and
     Y.\,-Z. Zhang, {J. Phys.\/} {\bf A 36} (2003), 7649.
\bibitem{Yan07} W.\,-L. Yang, Y.\,-Z. Zhang and X. Liu, {\it Phys.
     Lett.\/} {\bf B 641} (2006), 329; {\it J. Math. Phys.\/}
     {\bf 48} (2007), 053514.
\bibitem{Yan08} W.\,-L. Yang and Y.\,-Z. Zhang,
     {\it Nucl. Phys.\/} {\bf B 800}
      (2008), 527.
\bibitem{Yan08-1} W.\,-L. Yang, Y.\,-Z. Zhang, {\it Phys. Rev.\/}
     {\bf D 78} (2008), 106004, {\tt arXiv:0806.2477}.
\bibitem{Fra00} L. Frappat, P. Sorba and A. Sciarrino,
     Dictionary on Lie algebras and superalgebras, Academic Press,
     New York, 2000.
\bibitem{Kos74} B. Kostant, {\it Lecture Notes in Math} {\bf 466}
(1974), 101.













\bibitem{Kho91} S. M. Khoroshkin and V. N. Tolstoy, {\it Commun.
Math. Phys.} {\bf B 141} (1991), 599.






























\bibitem{Lep84} J. Lepowsky and R.\,L. Wilson, {\it Invent. Math.}
     {\bf 77} (1984), 199; {\it Invent. Math.} {\bf 79} (1985),
     417; J. Lepowsky and M. Primc, {\it Contemp. Math.} {\bf 46}
     (1985).
\bibitem{Pri02} M. Primc, {\tt math.QA/0205262}.
\bibitem{Zha05} Y.\,-Z. Zhang, X. Liu and W.\,-L. Yang,
     {\it Nucl. Phys.\/} {\bf B 704} (2005), 510.












\end{thebibliography}
\end{document}